\documentclass[a4paper,12pt,reqno]{amsart}

\usepackage[centertags]{amsmath}
\usepackage{amsfonts}
\usepackage{euscript}
\usepackage{amssymb}
\usepackage{amsthm}
\usepackage{newlfont}
\usepackage{stmaryrd}
\usepackage{mathrsfs}

\DeclareRobustCommand{\SkipTocEntry}[4]{}

\theoremstyle{plain}

\theoremstyle{definition}

\theoremstyle{remark}

\numberwithin{equation}{section}

\let\al=\alpha \let\be=\beta \let\de=\delta \let\ep=\epsilon
   
\let\ka=\kappa \let\la=\lambda \let\om=\omega 
\let\si=\sigma

\let\De=\Delta  \let\La=\Lambda \let\Om=\Omega

\newcommand{\caB}{{\mathcal B}}

\newcommand{\caH}{{\mathcal H}}

\newcommand{\caL}{{\mathcal L}}

\newcommand{\caP}{{\mathcal P}}
\newcommand{\caQ}{{\mathcal Q}}

\newcommand{\caZ}{{\mathcal Z}}

\newcommand{\bbC}{{\mathbb C}}

\newcommand{\bbE}{{\mathbb E}}

\newcommand{\bbP}{{\mathbb P}}

\newcommand{\bbR}{{\mathbb R}}

\newcommand{\bbZ}{{\mathbb Z}}

\newcommand{\opunit}{\text{1}\kern-0.22em\text{l}}

\newcommand{\frS}{{\mathfrak S}}

\newcommand{\bsP}{{\boldsymbol P}}

\DeclareMathAlphabet{\mathpzc}{OT1}{pzc}{m}{it}

\newcommand{\rel}{\,|\,}

\newcommand{\id}{\textrm{d}}
\newcommand{\can}{\text{can}}

\DeclareMathOperator{\Tr}{Tr}
\DeclareMathOperator{\tr}{tr}

\begin{document}

\begin{center}
\noindent {\Large {\bf A selection of nonequilibrium issues}}

\vspace{18pt} {\bf Christian Maes},\footnote{Instituut voor
Theoretische Fysica, K.~U.~Leuven, Belgium.\\ email: {\tt Christian.Maes@fys.kuleuven.be}} {\bf
Karel Neto\v{c}n\'{y}}\footnote{Institute of Physics AS CR, Prague, Czech Republic.\\ email: {\tt
netocny@fzu.cz}} and {\bf Bidzina Shergelashvili}\footnotemark[1]$^,$\footnote{On leave from
Georgian National Astrophysical Observatory, Kazbegi ave.~2a, 0160 Tbilisi, Georgia.}

\vspace{10mm}
{\sl Lecture notes from}\\
\vspace{3mm}
{\large {\sl the 5th Prague Summer School on\\
Mathematical Statistical Mechanics (2006)}}
\end{center}

\vspace{100pt} \footnotesize \noindent {\bf Abstract: } We give a
pedagogical introduction to a selection of recently discussed topics in nonequilibrium statistical
mechanics, concentrating mostly on formal structures and on general principles. Part I contains an
overview of the formalism of lattice gases that we use to explain various symmetries and
inequalities generally valid for nonequilibrium systems, including the fluctuation symmetry,
Jarzynski equality, and the direction of currents. In Part II we concentrate on the macroscopic
state and how entropy provides a bridge between microscopic dynamics and macroscopic
irreversibility; included is a construction of quantum macroscopic states and a result on the
equivalence of ensembles.
\normalsize
\newpage
\tableofcontents

\newpage
\vspace{15mm}
\begin{center}
{\large{\bf Part I.\\ Fluctuations in stochastic lattice gases}}
\end{center}
\vspace{3mm}

\addtocontents{toc}{\textbf{Part I. Fluctuations in stochastic lattice gases}}

\section{Introduction}

A good way to learn about possible constructions of nonequilibrium statistical mechanics probably
proceeds via the study of simple model systems.  Traditionally, the so called stochastic lattice
gases are playing there a prime role. A very early example is the Ehrenfest model. Not only was it
important as one of the many urn models illustrating strategies and results from probability
theory and from statistics, but it remains instrumental in learning about relaxation and about
detailed balance, see e.g.~\cite{K} where Mark Kac does not seem to hesitate in calling the
Ehrenfest model one of the most important models in all of physics. We will
encounter that Ehrenfest model in the first section.\\

Over the last decades, many other lattice gas models have been invented.  Often they appear
attractive because they obey simple updating rules and they are rather easy to visualize and to
simulate (at least today).  Yet, their behavior is rich, including sometimes clear examples of
emergent behavior. The latter refers to the organization of robust structures or patterns, of
critical behavior, and of phase transitions, which result from some collective or cooperative
behavior between the many interacting components. They have appeared in interdisciplinary
contexts, varying from models of traffic, to models of turbulence or to models for the spreading
of infections, in computer science, in economy etc. On the more mathematics side, we have here an
interesting ground for exploring and extending the theory of spatially extended Markov processes.
The role of the spatial architecture of the processes has recently been more in the center of
attention, e.g.~in discussions of processes on random graphs, small worlds etc. In recent
versions, the architecture (or graph) also undergoes a dynamics, in interaction with the
particles.
 Clearly these lattice gas models have proven their use already. A
few books where mathematical and statistical mechanical introductions are found to the theory of
lattice gases are \cite{kipnislandim,liggett,spohn}.\\

Many good references link stochastic lattice gases with fundamental problems in physics, be it in
the context of turbulence or in the derivation of hydrodynamic equations. In the present notes, we
bring together a number of recent results in the construction of nonequilibrium statistical
mechanics, as they appear for some simple stochastic lattice gas.  The emphasis will be mostly on
formal relations, from which both mathematical and physics treatments can find inspirations.  We
hope that this provides some step in a wider understanding of nonequilibrium issues in other and
more realistic models.  Indeed, one should remain aware that stochastic lattice gases are often
only effective tools. They are Markovian from the start and their transition rates depend on some
{\it ad hoc} choices. They are stochastic and there is no specification of a larger
environment.\\

After a short reminder of aspects of a Markov dynamics for one particle, we introduce the main
models in Section 3.  That is continued in Section 4 where the steady state is further specified.
In all that we work with a finite one-dimensional system on which there is a particle hopping in
the bulk of the system and a birth and death process at the two boundary sites. We consider a
time-dependent version of the dynamics in Section 5. The main tool is provided by a Lagrangian
set-up in which a Girsanov formula specifies the action (Section 6).  Section 7 gives the main
fluctuation relations in the form of a Jarzynski identity (relating the irreversible work with the
change in free energy) and of a fluctuation symmetry in the particle current (a so called steady
state fluctuation theorem).  While the results are mainly well-known we are not aware of a similar
{\it unifying} presentation in the literature.  There remain however very many nonequilibrium
issues which are not treated in these notes.  Some remarks are devoted to them in the final
section.

\section{One walker}

Consider the set $K=\{0,\ldots,N\}$ and the discrete time Markov chain with transition
probabilities $p(x,x') = x/N$ if $x'=x-1$ and $p(x,x') = 1- x/N$ if $x'=x+1$ for $x,x'\in K$. One
interpretation is to think of the state $x \in K$ as the number of particles in one of two
vessels.  The total number of particles over the two vessels is fixed equal to $N$.  At each
discrete time moment, one of the $N$ particles  is randomly selected and moved to the other vessel
from where it was.
That model is known as the Ehrenfest model (or dog-flea model).\\
 As a mathematical object we have here
an irreducible (but not aperiodic!) Markov chain $(x_n)$ that satisfies the condition of detailed
balance with respect to the stationary measure
\[
\rho(x) =2^{-N}\,\frac{N!}{x!(N-x)!}\,,\qquad x\in K
\]
That condition of detailed balance
\[
p(x,x')\;\rho(x) = p(x',x)\;\rho(x')
\]
expresses the time-reversibility of the stationary process. Indeed, let $\bsP_\rho$ denote the
stationary process on
$K^{\bbZ}$ and define $y_n = x_{-n}$.  The process $(y_n)$ is
Markovian with stationary law $\rho$. Its law is denoted by
$\bsP_\rho\Theta$ where $\Theta$ stands for time-reversal.  We
show that $\bsP_\rho\Theta =\bsP_\rho$ as a consequence of the condition of detailed balance. The
basic observation is that the transition probability for the process $(y_n)$ is via Bayes' formula
\[
q(y,y') = \mbox{Prob}[y_{n+1}=y' \rel y_{n}=y] = p(y',y)\,\frac{\rho(y')}{\rho(y)} = p(y,y')
\]
Therefore, the condition of detailed balance is equivalent with
the time-reversibility.\\
There is an easy way to generalize the above set-up.  Let us first make the step to continuous
time.  We are now speaking about rates
$c(x,y) \geq 0$ (or, transition probabilities per unit time) for
the transition $x\rightarrow y$. If we assume that \[ c(x,y) = a(x,y) e^{[V(x) - V(y)]/2} \] where
$a(x,y)=a(y,x)$ is symmetric, then still
\begin{equation}\label{dba}
\frac{c(x,y)}{c(y,x)} = e^{[V(x) - V(y)]}
\end{equation}
and $\rho(x) \propto \exp[-V(x)]$ is a reversible measure.\\
A new interpretation arises when thinking of the set
$\{0,1,\ldots,N\}$ as the sites of a lattice interval, with the
usual nearest neighbor connections.  The rates $c(x,y)$ could be taken non-zero only if $y=x\pm 1$
in which case we have a nearest neighbor walk.  The condition of detailed balance \eqref{dba}
assures that the walker will not drift; there is a potential landscape
$V(x)$, $x=0,\ldots,N$ which
could be periodically repeated to cover all of $\bbZ$ if wished.

There are ways to break detailed balance. One could for example insert a non-zero transition rate
for moving between the states (sites) $0 \leftrightarrow N$, and then take $c(x,x+1) = p$,
$c(x,x-1) = q\neq p$, $N + 1=0$.  In that case, say with $p> q$ there is a drift that the particle
moves more $x\rightarrow x+1$ than
$x\rightarrow x-1$; there appears a net current.\\
More generally, we can think of parameterizing the rates via
\[
c(x,y) = a(x,y)\, e^{[V(x) - V(y)]/2}\,e^{s(x,y)/2}
 \]
 where $s(x,y) = -s(y,x)$ would be antisymmetric.  It turns out
 that this term $s(x,y)$ has often an interesting physical
 interpretation.  In what follows we will see it related to the
 entropy production.  The entropy production is a physical notion
 that has arisen within irreversible thermodynamics, see e.g \cite{GM}.
 It goes well with considerations close to equilibrium.  The influence of the
 time-symmetric factor $a(x,y)$ is less understood.\\

 The following sections will study some of the aspects above for
 multi-particle models. We now have a (possibly variable)
 number of particles and they
 move on the lattice following certain hopping rules.  The effect
 of having many particles can result in (simpler) hydrodynamic
 behavior for macroscopic variables such as the density profile,
 but we will concentrate on the fluctuations instead.

\section{Stochastic lattice gases}
We start with a description of what is typically involved in stochastic lattice gases.  We do not
give the most general definitions but we specify to one special class.

\subsection{States}
By a lattice gas we understand a collection of particles whose positions are confined to the sites
of a lattice. In some models the particles still have a momentum, most often with a finite number
of possible values. Or, the particles can have extra decorations such as color or spin.  Here we
do not consider that.\footnote{An important ingredient of the Hamiltonian (or symplectic)
structure is thus lost.  In particular the kinematical time-reversal that would normally change
the sign of the velocities is absent.} The system thus consists of identical particles that can
jump from site to site on the given architecture. The states of the system are assignments to each
site of the number of particles. To be specific we consider the finite linear chain $\Lambda_N =
\{-N,-N+1,\ldots,0,1,\ldots,N-1,N\}$.  The endpoints $i=\pm N$ in
$\Lambda_N$ will play a special role in what follows; we call them
the boundary of the system while the other sites are in the bulk.
Two sites $i,j$ are nearest neighbors when $j=i\pm 1$.\\
 We allow
at most one particle per site $i$. We say that site $i$ can be vacant or occupied. The state space
(or the configuration space) is the finite set $K = \{0,1\}^{\Lambda_N}$. Elements of $K$ are
denoted by $\eta,\eta',\xi,\ldots$ and we write $\eta(i)\in
\{0,1\}$ for the occupation at site $i\in \Lambda_N$.

\subsection{Energy, entropy and particle number}
One imagines a function $H$ typically referred to as the Hamiltonian of the system,\footnote{It is
of course not a Hamiltonian in the strict sense of analytical mechanics.  The word Hamiltonian
refers here more to the quantum world where one considers for example the hopping of electrons in
a crystal structure.  A mathematically precise correspondence, also for the dynamical properties,
can often be achieved via the so called weak coupling limit or within the framework of Fermi's
Golden Rule.} that measures the energy of the state $\eta$. There is a great freedom of choice and
all depends on the context or on the specific purpose.  It does not hurt however to suppose
something specific, say an energy
function consisting of two terms:%
\begin{equation}\label{hamilt}%
H(\eta)=-B\sum_{i=-N}^{N} \eta(i) - \kappa \sum_{i=-N}
^{N-1} \eta(i)\,\eta(i+1),%
\end{equation}%
where $B$ and $\kappa$ are some constants. The first term contributes an energy $-B$ per particle
being present in the system and the second term takes into account some form of nearest neighbor
interaction related to the relative concentration
of particles on neighboring sites.\\

Speaking of energy reminds us of its conservation law. We can indeed imagine that our system is in
thermal contact with a very large heat bath at inverse temperature $\beta$ (Boltzmann's constant
is set equal to one), and for which all relevant changes are determined by the transitions in the
system.  In particular, every change $H(\eta') -
 H(\eta)$ in energy of the system is accompanied with the opposite change
 $\Delta E(\eta,\eta')= -(H(\eta') -
 H(\eta))$ of
energy in the bath.\footnote{In a more microscopic set-up, including a description of the degrees
of freedom in the heat bath, one would  need to specify a more exact decomposition of the total
energy into the system part and the part that belongs to the reservoir.  There would also be
interaction terms, the coupling, that contain both system and reservoir variables.  Obviously,
some convention is then needed of what is system and what is reservoir variable.} Imagining that
the energy change of the reservoir is thermodynamically reversible, we associate to it a change of
entropy in the reservoir equal to \[ \Delta S_{\text{res}} =
\beta\Delta E(\eta,\eta') =  -\beta (H(\eta') -
 H(\eta))
 \]
In other words, every change $\eta \rightarrow \eta'$ in the system's configuration entails an
entropy flux, that is $\beta$ times the heat dissipated in the thermal reservoir.\footnote{On a
scale where one supposes that the $\eta$ give the full microscopic description of the system,
there is no associated change of entropy in the system.  The total change of entropy (also called,
the entropy production) is then also equal to $-\beta (H(\eta') -
 H(\eta))$.  Most of the time however, there is a further {\it lower} level of
  description of the system variables with an associated degeneracy.}\\
For equilibrium purposes with just one heat bath, the relevant thermodynamic potential is the
Helmholtz free energy. Its statistical mechanical version is
\[
F = -\frac 1{\beta} \log Z\,,\qquad Z=\sum_{\eta\in K} e^{-\beta H(\eta)}
\]
Observe that if we change some parameter in $H$, e.g.~the coupling coefficient $\kappa$ in
\eqref{hamilt} (for fixed temperature), then the change in free energy $F=F(\kappa)$ equals the
expected change in energy:
\[
\frac{\id F}{\id\kappa} = \Bigl\langle \frac{\id H}{\id\kappa}\Bigr\rangle,\qquad
H=H_\kappa
\] where
\[ \langle g\rangle = \frac 1{Z}\sum_{\eta\in K} g(\eta)\,
e^{-\beta H(\eta)}
\] is
the thermal expectation.\\

Another important observable is the particle number.  We write
\[
{\cal N}_{[j,k]}(\eta) =\sum_{i=j}^k\eta(i)
\]
for the total number of particles in the lattice interval
$[j,k]\cap\Lambda_N$, $-N\leq j\leq k\leq N$. By construction, here
we have that the particle numbers ${\cal N}_{[j,k]}\leq 2N+1$ are {\it a priori} uniformly
bounded.  The total number of particles
is denoted by $\cal N = {\cal N}_{[-N,N]}$.\\
Making the correspondence with a gas again makes us think of a conservation law, now of the total
number of particles. In what follows, we imagine that the system is also in contact with a
particle reservoir at its boundary. Through the endpoints $i=\pm N$ particles can enter or leave
the system. We can also speak of a birth or a death of a particle at these sites. In Section
\ref{currents} we will introduce the particle currents.  As
particles can carry energy (see e.g.\ the first term in
\eqref{hamilt}), the flow of particles in and out of the system
can also contribute to the change of energy in the reservoir, and
hence to changes in entropy.\\
The equilibrium ensemble that allows both the exchange of energy and of particles is the
grand-canonical one.  It gives probabilities
\begin{equation} \label{gibbs}%
\mathbb{P}^{\beta,a} [\eta]=\frac 1{{\cal Z}}\,e^{a\,\sum\eta(i)}\,e^{-\beta H(\eta)}%
\end{equation}%
where ${\cal Z}={\cal Z}(a,\beta,N)$ is a normalization factor. The constant $a$ is called the
chemical potential and in equilibrium it refers to and it is determined by the concentration of
particles in the (imagined very large) environment.

\subsection{Dynamics}
The dynamics is given by a continuous time Markov process on $K$. We distinguish two modes of
updating:
\begin{itemize}
\item A particle can jump (or hop) to nearest neighbor sites. That
is a diffusion mechanism.  We will not add external fields to the dynamics not to impose a bulk
drift or bias;
 \item Particles can leave or enter the system at the boundary.  That is a
reaction mechanism.  The system will be boundary driven.
\end{itemize}
We introduce some further notation to formalize the dynamics. As we only consider symmetric
hopping, it is useful to introduce the transformation
\begin{equation}
\eta ^{i,j} (k) =
\begin{cases}
\eta (k) & \text{if } k\neq i,\, k\neq j;
\\
\eta (i) & \text{if } k= j;
\\
\eta (j) & \text{if } k= i
\end{cases}
\end{equation}
That defines the configuration obtained from $\eta$ after switching the occupation of the sites
$i,j$. We allow only the hopping of particles to neighboring sites $j = i \pm 1$. The rate of the
transition due to that diffusion mechanism is taken as
\begin{equation}\label{cij}%
C(i,j,\eta)=\exp \Bigl[ -\frac{\beta}{2} (H(\eta^{i,j})
-H(\eta))\Bigr]\,,\quad |i-j|=1%
\end{equation}%
where $H(\eta^{i,j})$ is the energy function after the transition and $H(\eta)$ is that
corresponding to the initial
configuration.\\

 Analogously, we define the rate of birth and death
$\eta \rightarrow
\eta ^{i}$ of the particles as%
\begin{equation}\label{ci}%
C(i,\eta)=e^{-a_{i}\eta(i)}\exp\Bigl[ -\frac{\beta}{2} (H(\eta^{i}) - H(\eta))\Bigr]
\end{equation}%
where $H(\eta^{i})$  is the energy function after the transition to $\eta^i$, the new
configuration
after the birth or the death of a particle at site $i$:%
\begin{equation}
\eta^{i}(k) =
\begin{cases}
1- \eta(k) & \text{if } k=i
\\
\eta(k) & \text{if } k\neq i
\end{cases}
\end{equation}
To be definite we take births and deaths only at the boundary
sites $i=-N,N$.\\

The physical interpretation of the dynamics is quite simple. Think of a one-dimensional channel in
which particles diffuse while they enter or leave the system at its boundary.  A biophysical
realization seems  to be found in the physics of ion channels connecting the inside and the
outside of a living cell.  The channel is a sort of opening or gate in the cell's membrane through
which charged particles can move from higher to lower concentration, or following the gradient in
electric potential etc.  Here the relevant parameters are the values $a_{\pm N}$ which in fact
represent the (different) chemical potentials of the
two reservoirs at the outer edges.\\

 With these definitions the
Master equation governing the temporal behavior of probability
measures on $K$ is given by%
\begin{equation}\label{master}
\begin{split}
\frac{\id}{\id t} \mathbb{P}_{t}(\eta) &= \sum _{i=1}^{N-1}[C(i,i+1, \eta^{i,i+1})
\mathbb{P}_{t}(\eta^{i,i+1}) - C(i,i+1, \eta)
\mathbb{P}_{t}(\eta)]
\\
&\hspace{5mm}+ C(-N,\eta^{-N}) \mathbb{P}_{t}(\eta^{-N}) - C(-N,\eta) \mathbb{P}_{t}(\eta)
\\
&\hspace{5mm}+ C(N,\eta^{N}) \mathbb{P}_{t}(\eta^{N}) - C(N,\eta) \mathbb{P}_{t}(\eta)
\end{split}
\end{equation}
That equation shows how the probability to find a given configuration in the system evolves in
time.  Alternatively, the generator $L$ is given by
\begin{equation}\label{mastexp}%
\frac{\id}{\id t} \langle f(\eta _t)\rangle=\langle L f(\eta _t)\rangle%
\end{equation}%
for functions (``observables'') $f$ on $K$, and where $\langle
\cdot \rangle$ takes the expectation over the Markov process,
including some (as yet unspecified) initial distribution. Explicitly,
\begin{multline}\label{kineq}
L f(\eta) = \sum _{i=1}^N\,C(i,i+1, \eta) [f(\eta^{i,i+1}) - f(\eta)]
\\
+ C(-N,\eta)[f(\eta^{-N}) -f(\eta)] + C(N,\eta)[f(\eta^{N}) - f(\eta)]
\end{multline}
Let us make a simple exercise by plugging in $f(\eta) = \eta(i)$ for some fixed $i$ and by taking
$\beta=0$ in
\eqref{cij}--\eqref{ci}. The corresponding evolution equation is
\[
\frac{\id}{\id t}\langle \eta_t(i)\rangle = \langle \eta_t(i-1) +
\eta_t(i+1) - 2\eta_t(i)\rangle
\]
when $i\neq -N,N$, while for $i=\pm N$,
\[
\frac{\id}{\id t}\langle \eta_t(i)\rangle = \langle \eta_t(i\mp 1) -
\eta_t(i) + e^{-a_{i}\eta_{t}(i)}[1-2\eta_t(i)]\rangle
\]
Apparently, these equations are closed in the density variables
$\langle \eta_t(i)\rangle$, $i\in \Lambda_N$.  In particular,
putting their left-hand sides equal to zero, we get the stationary value $\langle \eta(i)\rangle =
Ci + D$ for some constants $C$ and
$D$ that depend on $N$ and on the values $a_{\pm N}$.  One checks
that $a_{-N}=a_N=a$ if and only if $C=0$, $D=1/(1+ \exp(-a))$.  When
$C\neq 0$, then there is a linear density profile with slope $\sim
1/N$.  Obviously, when repeating that calculation for $\beta\neq 0$, we run into a difficulty: the
equation for the $\langle
\eta_t(i) \rangle $ is no longer closed but there is a coupling
with higher order correlation functions such as $\langle
\eta_t(i)\,\eta_t(i+1)\rangle$. That feature is very generally true
and it implies that we cannot simply solve the equations.\footnote{The problem appears in all
nontrivial dynamics for many particle systems. The resulting hierarchy of equations is sometimes
referred to as the BBGKY-hierarchy, referring in particular to the hierarchy of equations that
appear in kinetic gas theory for the various particle distribution functions. The study of
possible ways of closing the hierarchy is a major concern in nonequilibrium physics.} The
stationary distribution is in general only implicitly known, as solution of the (time-independent)
Master equation~\eqref{master} with the left-hand side set zero.

\subsection{Path-space measure}

 One has to remember that a Markov
process is a special probability distribution on paths.  In our case, we have piecewise-constant
paths. A path $\omega$ over the time-interval $[0,\tau]$ starts from an initial configuration
$\eta_0$ after which it changes into $\eta_{t_1},
\eta_{t_2},\ldots$ at random times $t_1, t_2,\ldots$ To be more
precise we must add what is the configuration at the jump times as well.  That is just a
convention, and we take it that
$\eta_{t_{k-1}} = \eta_{t_k-}$ and $\eta_{t_k} = \eta_{t_k+}$, or, the
step-function is continuous
from the right.\\
An important transformation on path-space concerns the so called time-reversal $\Theta$ in which
$(\Theta\omega)_t  =
\omega_{\tau-t}$, up to irrelevant modifications at the jump times
making $\Theta\omega$ again right-continuous.

 The random times are called the jump times of
the process. The Markov process assigns a probability law to these times and to the corresponding
transitions. There are two ingredients: the waiting time and the transition step. The waiting time
is exponentially distributed with a weight $\lambda(\eta)$ that depends on the present
configuration $\eta$. That waiting time is directly (and inversely) related to the escape rate
\[
\lambda(\eta) = \sum_{\eta'}
 W (\eta \rightarrow \eta')
 \]
 We
will use the notation $W(\eta\rightarrow \eta')$ when
indicating one allowed but general transition rate.\\
 The second
ingredient sits in the transition rates as we have them introduced before.  When the waiting time
is over, a new configuration is chosen so that for time $t\downarrow 0$,
\[
\mbox{Prob}[\eta_t=\eta' \rel \eta_0=\eta] = (1 - \la(\eta)\, t)\, \delta_{\eta,\eta'} +
W(\eta\rightarrow \eta')\, t + o(t)
\]
A more explicit realization of that path-space measure goes via Girsanov's formula, see Section
\ref{gir}.

\section{Steady state}\label{stst}
\subsection{Detailed balance}
One observes from the definition (\ref{cij}) that:
\begin{equation}\label{detail1}
    \frac{C(i,j,\eta)}{C(i,j,\eta^{i,j})}=
    \frac{\exp \left [ -\beta H\left (
\eta^{i,j}\right ) \right ]}{\exp \left [ -\beta H\left (
\eta\right ) \right ]}= \frac{\mathbb{P}^{\beta,a}
[\eta^{i,j}]}{\mathbb{P}^{\beta,a} [\eta]}
\end{equation}
where we have inserted the ratio of probabilities according to
\eqref{gibbs}. That is verified for all values $a$.  Furthermore,
with the definition (\ref{ci}) we have
\begin{equation}\label{detail2start}%
\frac{C(i,\eta)}{C(i,\eta^{i})}= \frac{\exp [-a_{i} \eta(i)]}{\exp
[-a_{i} (1-\eta(i))]}\frac{\exp \left [ -\beta H\left (
\eta^{i}\right ) \right ]}{\exp \left [ -\beta H\left ( \eta
\right ) \right ]}\,,\quad i=\pm N%
\end{equation}%
If $a_{-N}=a_{N}=a$ then
\[
\frac{\exp [-a_{i} \eta(i)]}{\exp [-a_{i} (1-\eta(i))]}=
\frac{e^{a(1-\eta(i))}}{e^{a\eta(i)}}
\]
Comparing with formula \eqref{gibbs}, still  for $i=\pm N$ and for
$a_{-N}=a_{N}=a$,
\begin{equation}\label{detail2}%
\frac{C(i,\eta)}{C(i,\eta^{i})}= \frac{\mathbb{P}^{\beta,a}
[\eta^{i}]}{\mathbb{P}^{\beta,a} [\eta]}
\end{equation}%
Summarizing, when the particle reservoirs left and right have equal concentration, then the system
dynamics satisfies the condition of detailed balance
\begin{equation}\label{detailbalance}%
\frac{W(\eta\rightarrow \eta')}{W(\eta' \rightarrow
\eta)}=\frac{\mathbb{P}^{\beta,a} [\eta']}{\mathbb{P}^{\beta,a}
[\eta]}
\end{equation}%
for all allowed transitions $\eta \rightarrow \eta'$ and corresponding transition rates
$W(\eta\rightarrow \eta')$. Under that same condition $a_{-N}=a_{N}=a$ we thus have that
\eqref{gibbs} is a reversible stationary measure. The
corresponding process is the
steady state for equilibrium conditions.\\

Observe that if we consider unequal rates at the boundaries $a_1
\neq a_N$ then we could still try
\begin{equation}%
\mathbb{P}^{\beta,a_1,a_N}(\eta)=\frac 1{Z}\exp [-\beta
H(\eta)]\exp
[a_1 \eta(1)+a_N\eta(N)]%
\end{equation}%
as a candidate stationary distribution. In that case the analogue of \eqref{detail2} is still
verified.
 Yet, the condition  \eqref{detail1} fails.
\subsection{Nonequilibrium model}\label{currents}
Now comes the question what happens when $a_1\neq a_N$. Let us first consider the left boundary of
the system, for which we can write, see \eqref{detail2start},
\begin{equation}%
\label{aneqa1}%
\frac{C(-N,\eta)}{C(-N,\eta^{-N})}=e^{-\beta
[H(\eta^{-N})-H(\eta)]-a_{-N} J_{\ell} (\eta, \eta^{-N})}%
\end{equation}%
where $J_{\ell}(\eta, \eta^{-N})=1$ when the
 particle leaves the system via the site $-N$, i.e., $\eta(-N)=1$,
and $J_{\ell}(\eta, \eta^{-N})=-1$ when a new particle enters, i.e., $\eta(-N)=0$.  That is an
antisymmetric current of particles, taking positive when the particles leave the system. In the
same way we define the current $J_{r}(\eta, \eta') = 1$ when $\eta(N)=1$, $\eta'=\eta^N$ and
$J_{\ell}(\eta, \eta^{'})=-1$
when $\eta(N)=0$, $\eta'=\eta^N$.  The currents are zero otherwise.\\

Taking all transitions together, we have%
\begin{equation}\label{ldb}%
\frac{W(\eta\rightarrow \eta')}{W(\eta' \rightarrow
\eta))}=e^{-\beta[H(\eta')-H(\eta)]-a_{-N} J_{\ell}(\eta,
\eta') - a_N J_r(\eta, \eta')}%
\end{equation}%
One recognizes the change of entropy in the environment: \[ S(\eta,\eta') = \beta\Delta
E(\eta,\eta') - \mu_\ell \Delta {\cal N}_\ell(\eta,\eta') - \mu_r \Delta {\cal N}_r(\eta,\eta')
\]
where $\mu_\ell= a_{-N}$ respectively $\mu_r=a_{N}$ are the chemical potentials (up to some factor
$\beta$ that we have ignored) of the particle reservoirs left and right, and $J_\ell =
\Delta {\cal N}_\ell$, $J_r = \Delta {\cal N}_r$ are the changes in
particle number in the {\it left}, respectively {\it right} particle reservoir. The form
\eqref{ldb} or
\[
\frac{W(\eta\rightarrow \eta')}{W(\eta' \rightarrow \eta)}=
e^{S(\eta,\eta')}
\]
 is known as that of local detailed balance.\footnote{Remark that in \eqref{ldb}
 a possible time-symmetric prefactor to the rates
 \eqref{ci} or \eqref{cij} will never appear;
  there is only the part that is antisymmetric under $\eta\leftrightarrow \eta'$.
  The fact that the entropy production appears as the source term
  of the breaking of time-reversal symmetry, is no accident but it is related to more general
  considerations that here are simply applied in order to obtain
   a reasonable physical interpretation of
  our effective dynamics, see e.g.~\cite{poincare,mn}.}\\

The currents $J_\ell$ and $J_r$ appear in the conservation law for the particle number. The sum of
these currents equals
 the number of particles that leave the system,
\begin{equation} \label{conserlow}%
J_\ell (\eta, \eta')+J_r (\eta,\eta')={\cal N}(\eta)-{\cal N}(\eta')%
\end{equation}%
or%
\begin{equation}%
a_{-N}J_\ell (\eta, \eta')+a_N J_r (\eta,
\eta')=(a_{-N}-a_N)J_\ell+a_N ({\cal N}(\eta)-
{\cal N}(\eta'))%
\end{equation}%
> From now on, we write $a_N = a$, $a_{-N} = a + \delta$ so that
\[
\frac{W(\eta\rightarrow \eta')}{W(\eta' \rightarrow
\eta)}=\frac{\mathbb{P}^{\beta,a} [\eta']}{\mathbb{P}^{\beta,a}
[\eta]}\, e^{-\de J_\ell(\eta,\eta')}
\]
The parameter $\delta$ thus measures some distance to the equilibrium situation, and enables the
tentative terminology of
{\it close} versus {\it far} from equilibrium.\\

As above we define the bulk currents $J_i(\eta,\eta')$ to be $+1$ if in the transition
$\eta\rightarrow \eta'$ a particle moves over the bond $i\rightarrow i+1$, and equal to $-1$ if a
particle moves
$i \leftarrow i+1$.  More generally,\footnote{In fact and
throughout we call current what is more like a time-integrated current, or a change of particle
number.} we consider a path $\omega = (\eta_t)_{t=0}^\tau$ and currents $J_i(\omega)$,
$i=-N,\ldots,N,$ defined by \[ J_i(\omega) = J_i(\eta_0,\eta_{t_1}) + J_i(\eta_{t_1},\eta_{t_2}) +
\ldots + J_i(\eta_{t_{n-1}},\eta_{\tau})
\]
In particular, $J_r = J_N$ and for $i\leq k$,
\begin{align}\label{bc}
J_{i} (\omega) - J_{k}  (\omega) &= {\cal
N}_{[i+1,k]}(\eta_\tau)- {\cal N}_{[i+1,k]}(\eta_0)\nonumber\\
J_\ell(\omega) + J_{-N}(\omega) &= \eta_0(-N) - \eta_\tau(-N)
\end{align}%
Observe that the currents $J_i$ are extensive in the time
$\tau$.\\

All of that is related to the process, be it transient or be it steady.  Except for the following
section however, we will be mostly interested in the steady state regime.  It is easy to verify
that we have here a unique stationary distribution $\rho$. It satisfies the time-independent
Master equation \eqref{master} (zero left-hand side).  Corresponding to $\rho$ there is then a
stationary process with distribution $\bsP_\rho$.  If we look at expectations in the stationary
process we write
$\langle \,\cdot\, \rangle_\rho$.\\
> From the conservation laws \eqref{conserlow} and \eqref{bc} we have
\[
\langle J_\ell \rangle_\rho = -\langle J_r \rangle_\rho = -
\langle J_i \rangle_\rho, \qquad i \in \Lambda_N
\]
There are alternative expressions for these expectations by using the dynamical equations
\eqref{mastexp}.  For example, for $i\neq N,-N$,
\[
\frac 1{\tau}\langle J_i \rangle_\rho = \langle C(i,i+1,\eta)
(\eta(i) - \eta(i+1)\rangle_\rho =\langle 1- \eta(N)[1+ e^{-a}]
\rangle_\rho
\]


\section{Time-dependent dynamics}\label{td}

\subsection{Modifications with respect to Section \ref{stst}}
Nonequilibrium conditions can be obtained in a variety of ways. The above gives a set-up for
boundary driven steady states. Another way of driving the system away from equilibrium is by
applying an external {\it bulk} field.   We consider here a modification which also frustrates the
system (as it cannot simply relax to equilibrium). We remain with the same states but the updating
becomes time-dependent.  The idea is that the values of parameters in the
Hamiltonian are changed while the dynamics enrolls.\\

We have a time-dependent Hamiltonian $H_t$ so that the transition rates $W_t(\eta\rightarrow
\eta')$ are also depending on the moment $t$ of the jump $\eta\rightarrow \eta'$. For example, the
rate for exchanging the occupation at sites $i$ and $j=i\pm 1$ is
\begin{equation}
C_t(i,j,\eta)= \exp \Bigl[ -\frac{\beta}{2} \left ( H_t
(\eta^{i,j}) - H_t(\eta)\right )\Bigr]\,,\qquad |i-j|=1%
\end{equation}
(compare with \eqref{cij}) depending on the time $t$.\\
There is no longer a very good sense in which we can speak about the stationary distribution.
Still we can consider for each $H_t$, $t\in [0,\tau]$, the corresponding Gibbs distribution
\begin{equation}\label{bg}
\rho_t(\eta) = \frac 1{\cal Z_t} \,e^{a \sum\eta(i)}\,e^{-\beta H_t(\eta)}%
\end{equation}%
where ${\cal Z_t}={\cal Z_t}(a,\beta,N)$ is now also time-dependent.  There is an associated  free
energy
\begin{equation}\label{fe}
{\cal A}_t = - \frac 1{\beta} \log \cal Z_t
\end{equation}

In the time-dependent case, we will only work with the dynamics for which $\delta=0$,
$a_{-N}=a_N=a$ fixed, i.e., there is just one particle reservoir and one heat bath
reservoir.\footnote{We can of course make a dynamics such that at every moment the distribution
$ \mathbb{P}_{t} =\rho_t$ exactly coincides with \eqref{bg}. We
could e.g. take  a transition rate $\bar W_t(\eta \rightarrow
\eta') = \rho_t(\eta')$.  One can think of it as admitting an
infinitely fast relaxation of the equilibrium process. Alternatively, one can think of an
ultra-slow time-dependence in
$H_t$ so that, before any change, the system has already relaxed
to the equilibrium distribution corresponding to the instantaneous value.}

\subsection{Work and heat}
When there is an energy exchange between system and reservoir, there is heat.  For a history
$\omega = (\eta_t)_0^\tau$ where the jumps in the configuration happen at times  $t_1,t_2,\ldots
t_n$, the total
heat $\cal Q$ transferred to the system is the sum of differences of energy:%
\begin{align}\label{heat}
{\cal Q}=H_{t_{1}}(\eta _{t_1})-H_{t_{1}}(\eta _{0})+H_{t_{2}}(\eta _{t_{2}})-H_{t_{2}}(\eta
_{t_{1}})+ \ldots &&\nonumber\\+H_{t_n}(\eta_\tau) - H_{t_n}(\eta_{t_{n-1}})&&
\end{align}%
On the other hand,  the work $\cal W$ performed upon the system is a sum of changes of the
Hamiltonian at
fixed configurations:%
\begin{align}\label{work}
{\cal W}=H_{\tau}(\eta _{\tau})-H_{t_{n}}(\eta _{\tau})+H_{t_{n}}(\eta
_{t_{n-1}})-H_{t_{n-1}}(\eta
_{t_{n-1}})+\ldots&&\nonumber\\
+H_{t_1}(\eta_0) - H_0(\eta_0)&&%
\end{align}%
Therefore, as an expression of the first law of
thermodynamics,%
\begin{equation}\label{1law}
\cal Q+ \cal W =H_\tau(\eta_\tau) - H_0(\eta_0) %
\end{equation}%
 is the total change of system energy between the initial
and the final configurations $\eta_0$ and $\eta_\tau$ in the path
$\omega$.

\section{Lagrangian set-up: Girsanov formula}\label{gir}

As we have seen in the course of our computation around and below
\eqref{kineq}, the evolution equations give a hierarchy of
equations for the various correlation functions of the stationary distribution.  Solving them is
like diagonalizing a large matrix and it is not even clear whether it would always permit us to
extract the  most relevant information. A more global characterization of the stationary
distribution is perhaps obtained by going to a space-time picture.  On that level the process is
space-time local and explicit.  The variables are the
histories or trajectories of the system. \\


Given two Markov processes on the same space $K$, we can consider two path-space measures $\bsP$
and $\bar{\bsP}$ with corresponding escape rates $\lambda$ and $\bar\lambda$, and transition rates
$W$ and $\bar W$.  We consider all paths on the interval $[0,\tau]$ and we assume that for all
$\eta$,
\[
\{\eta', W(\eta\rightarrow \eta') \neq 0\} = \{\eta', {\bar W}(\eta\rightarrow \eta') \neq 0\}
\]
We can then look for the density of $\bsP$ with respect to
$\bar{\bsP}$.  That density is a Radon-Nikodym derivative and can
be written down quite explicitly in the so called Girsanov formula:
\begin{equation}\label{girs}
\frac{\id\bsP}{\id\bar \bsP}(\omega) = \exp\Bigl[\int_0^\tau \,
\big(\lambda(\eta_t) -\bar \lambda(\eta_t)\big)\,\id t  + \sum_{t\leq
\tau}\log \frac{W(\eta_{t-}\rightarrow \eta_t)}{{\bar
W}(\eta_{t-}\rightarrow \eta_t)}\Bigr]
\end{equation}
when restricted to events that are measurable from the trajectory in $[0,\tau]$.  The last sum in
the exponential is over the jump times, as they appear in the path $\omega$.  We have assumed here
that the two processes $\bsP$ and $\bar \bsP$ start from the same configuration.  If they have
different initial distributions $\nu$ and $\bar \nu$, then a prefactor $\nu(\eta)/\bar\nu(\eta)$,
$\omega_0 =\eta,$ must be added to the right-hand side of the
Girsanov formula \eqref{girs}. The formula remains intact when the process is not
time-homogeneous.  One then adds the correct time-dependence to the escape and to the transition
rates.\footnote{The formula is also not really restricted to Markov processes, or to finite state
spaces.  The more useful way of considering that formula is as a generalization of the
Boltzmann-Gibbs formula \eqref{gibbs}, where the essential input is that one can make sense of
what is written in the exponential as a sum of quasi-local terms.  Here we can speak about the
action as a sum of a local Lagrangian. Indeed, if we write out the rates
$W(\eta\rightarrow \eta')$ of our stochastic lattice gas, and we
take $\bar\bsP$ say corresponding to $\beta=0$, $a_{-N}=a_N=0$ we obtain there a sum over
space-time of local interaction terms. Without trying to formalize the idea, see however
\cite{maes}, one can thus consider the stationary distribution to be the projection (or
restriction) of that space-time path-space measure to an equal time layer, see e.g.\ \cite{LMS}.
There is no {\it a priori} reason why that projected measure should inherit a
spatial locality, see e.g.\ \cite{mrv1}.}\\

Our application of the Girsanov formula will concern time-reversal.  If we have a distribution
$\bsP$ on paths, then its time-reversal $\bsP\Theta$ is obtained via
\[
\frac{\id\bsP\Theta}{\id\bsP^0}(\omega) =
\frac{\id\bsP}{\id\bsP^0}(\Theta\omega)
\]
for an arbitrary process $\bsP^0$ which is reversible.  The dependence on initial configurations
is again ignored, but it is essential in the  consideration of the time-reversal invariant process
$\bsP^0$.

\section{Fluctuation relations for the entropy production}
\subsection{Jarzynski equality}
Recall the set-up for the time-dependent dynamics in Section
\ref{td}. We take the case where $a_1=a_N=a$.
 The Jarzynksi
identity is a relation between the work $\cal W$ of \eqref{work} and the change in free energy
\eqref{fe}.  In the context of stochastic lattice gases, we get it as
\begin{equation}\label{jarzinski}%
\mathbb{E}_{\rho _0} ^a [e^{-\beta {\cal W}}]=e^{-\beta \Delta {\cal A}},%
\end{equation}%
The left-hand side is the expectation in the time-dependent dynamics with fixed chemical potential
(left and right) equal to
$a$ and with Hamiltonian $H_t$ at inverse temperature $\beta$,
started from equilibrium $\rho_0$ at time $t=0$.  The right-hand side contains the difference
\begin{equation}\label{dif}
\Delta \cal A = {\cal A}_\tau  - {\cal A}_0,%
\end{equation}%
of free energies\footnote{To make sure, there is no assumption that at any future time $t>0$
(including at time
$\tau$) the distribution should be the $\rho_t$ of \eqref{bg}.  We
are starting from equilibrium at time zero, but then the system is most likely away from
instantaneous equilibrium with respect to the Hamiltonian $H_t$.  Yet, the result of
\eqref{jarzinski} is a statement about equilibrium free energies.  We can measure these (and how
they possibly depend on some parameter) via some nonequilibrium procedure.}~\eqref{fe}.

\begin{proof}
 To prove (\ref{jarzinski}) we make a first application
of the Girsanov formula.  The two distributions correspond to our time-dependent process
$\bsP_{\rho_0}^a$ on the one hand and to the time-reversed process
$\bar{\bsP}_{\rho_\tau}^a\Theta$ on the other hand.
 By $\bar{\bsP}_{\rho_\tau}^a$ we mean the process started at
  time zero from the distribution $\rho_\tau$ and with time-reversed protocol, i.e.,
    the rates
  are $\bar W_t = W_{\tau -t}$.

We have therefore for a fixed path $\omega$ with jump times
$t_1,\ldots, t_n$ in the interval $[0,\tau]$, that
\begin{equation}\label{star}
\frac{\id\bsP_{\rho_0}^a}{\id\bar{\bsP}^a_{\rho_\tau}\Theta}(\omega)=
\frac{\rho_0(\omega_0)}{\rho_\tau(\omega _\tau)}\, \exp R(\omega)
\end{equation}
with, from \eqref{girs},
\begin{equation}\label{rs}
R=\frac{W_{t_1}(\eta _{0}\rightarrow \eta _{t_1})W_{t_2} (\eta _{t_1}\rightarrow \eta
_{t_2})\ldots W_{t_{n}} (\eta _{t_{n-1}}\rightarrow \eta _{\tau})} {W_{t_{n}} (\eta
_{\tau}\rightarrow \eta _{t_{n-1}})\ldots W_{t_2} (\eta _{t_2}\rightarrow \eta _{t_1})
{W_{t_1}(\eta _{t_1}\rightarrow \eta _{0})}}
\end{equation}%
By using the detailed balance relations
\[
\frac{W_t(\eta\rightarrow \eta')}{W_t(\eta'\rightarrow \eta)}=
\exp[-\beta(H_t(\eta')-H_t(\eta)) + a({\cal N}(\eta') - {\cal
N}(\eta))]
\]
and combining that with the expression \eqref{heat} for the heat, the ratio \eqref{rs} reduces to
\[
R(\omega) = e^{-\beta {\cal Q}}\, e^{a[{\cal N}(\eta_\tau) - {\cal N}(\eta_0)]}
\]
Hence, looking back at \eqref{star} and substituting \eqref{bg},
\eqref{1law} and \eqref{dif}, we have
\begin{align}\label{jare}
\log
\frac{\id\bsP_{\rho_0}^a}{\id\bar{\bsP}_{\rho_\tau}^a\Theta}(\omega)&=-\beta
{\cal Q}(\omega) + \log\frac{ {\cal Z}_\tau}{{\cal Z}_0} -
\beta[H_0(\eta_0) - H_\tau(\eta_\tau)]\nonumber\\
&= \beta[{\cal W}(\omega) - \Delta {\cal A}]
\end{align}

   The Jarzynski equality \eqref{jarzinski} is then an easy consequence of
  the normalization of path-space measures:
\begin{equation}%
\int \id\bsP_{\rho _0}^a (\omega)
\frac{\id\bar{\bsP}^a_{{\rho} _\tau}\Theta} {\id\bsP_{\rho _0}^a}(\omega)=1%
\end{equation}%
\end{proof}

For further background information on these relations between irreversible work and free energy
differences, one can check e.g.~\cite{Crooks,Jar,poincare}.

\subsection{The direction of particle current}
We come back to the time-homogeneous nonequilibrium process as we had it first in Section
\ref{currents}. Physically we expect that there will be a particle current flowing from higher to
lower concentration. To be specific, let us assume that $\delta \geq 0$, $a_{-N} \geq a_N,$ so
that the physical picture suggests that the mean particle current $\langle J_i\rangle_\rho \geq
0$. The question is how to actually see that. Remember that we do not know a thing about the
stationary distribution $\rho$ in general. Nevertheless the direction of the particle current will
easily follow within our set-up.\\

> From the Girsanov formula \eqref{girs} for $\bsP_\rho$ with respect to $\bsP_\rho\Theta$, both
started in the stationary distribution $\rho$, we have
\begin{equation}\label{conc}
\frac{\id\bsP_{\rho }}{\id\bsP_{\rho }\Theta }(\omega )=\frac{\rho
(\omega _{0})}{\rho (\omega _{\tau})}\exp \left[ -\beta \left(
H(\omega _{\tau})-H(\omega _{0})\right) +a\Delta {\cal N}-\delta J_{\ell}(\omega )\right],%
\end{equation}%
Again from the normalization we
have:%
\[
\int \id\bsP_{\rho} (\omega) \frac{\id\bsP_{\rho}\Theta}
{\id\bsP_{\rho}}(\omega)=1.%
\]
and hence, by concavity,
\[
\int \id\bsP_{\rho}
(\omega) \log \frac{\id\bsP_{\rho}\Theta} {\id\bsP_{\rho}}(\omega)\leq 0.%
\]
But, from \eqref{conc} and by stationarity
\begin{equation}%
0 \leq \int \id\bsP_{\rho}\,\log \frac{\id\bsP_{\rho}}
{\id\bsP_{\rho}\Theta}(\omega)=-\delta\langle J_\ell\rangle_\rho =
\delta\langle J_i\rangle_\rho
\end{equation}
 We conclude that
\begin{equation}%
\delta \langle J_i\rangle_\rho\geq 0%
\end{equation}%
which shows that the average direction of the particle current depends only on the sign of
$\delta$.   See \cite{mnv} for a very similar
analysis in the case of heat conduction.\\
To get a strict inequality $\langle J_i\rangle_\rho >0$ is also possible for $\delta >0$; it
suffices to see that there is a non-zero probability that the current $J_i$ as a function of the
path $\omega$ is not constant equal to zero even when
$\omega_0=\omega_\tau$.
\subsection{Fluctuation theorem}
The previous results were all a direct consequence of the normalization condition applied to the
Radon-Nikodym derivative
\eqref{girs} between two path-space measures. Here we go for a
result that is somewhat more detailed and concerns a symmetry in the fluctuations of the current.
We follow the method of \cite{poincare,maes,mrv2} but the
present model can also be treated via \cite{LS}.\\

We fix an $i=-N,\ldots,N$ and consider the current $J_i$ as function of the path over the interval
$[0,\tau]$. Define the generating function
$q(\lambda)$, $\lambda\in \bbR$ by
\begin{equation}\label{flucttheo}%
q(\lambda)=\lim _{\tau\uparrow +\infty} \frac{1}{\tau}\log
\langle e^{-\lambda J_{i}} \rangle_\rho%
\end{equation}%
The limit exists by the Perron-Frobenius theorem, and is independent of $i=-N,\ldots,N$ because of
\eqref{bc}.  The fluctuation symmetry is that
\begin{equation}\label{1c}
q(\lambda)=q(\delta-\lambda)%
\end{equation}%
 Before providing a proof
observe that $q(\lambda)$ is the Legendre transform of the rate functions of large deviations for
$J_{i}$.  The interested reader is referred to the literature on large deviations (and the
G\"artner-Ellis theorem in particular) for more details, see e.g.~\cite{DZ}. The idea is that
\begin{equation}\label{ld}
\bsP_\rho[ J_{i}  \simeq \tau j] \simeq e^{-\tau\,I(j)},\qquad
\tau\rightarrow +\infty
\end{equation}%
with%
\begin{equation}
I(j)=\inf _{\lambda} (-\lambda j-q(\lambda))%
\end{equation}%
Substituting the identity \eqref{1c}, we get
\begin{equation}%
I(j)= \inf _{\lambda} \big((-\delta + \lambda)(- j)
-q(\delta-\lambda)\big) - \delta \,j=-\delta\,j + I(-j)%
\end{equation}%
or
\begin{equation}%
I(j)-I(-j)= -\delta\,j %
\end{equation}%
That can be again translated to an identity for \eqref{ld}:
\begin{equation}\label{gc}
\frac{\bsP_\rho (J_{i} /\tau \simeq j)}{\bsP_\rho (J_{i}/\tau \simeq -j)}
\simeq e^{\tau\,\delta\,j}%
\end{equation}
The interpretation of that expression is that there exists a relation between the probabilities of
having a current $+j$ and
$-j$; the probability of having a particle current in the
direction opposite to the expected one ($\delta j
> 0$) is exponentially small with $\tau\rightarrow +\infty$.
  The formula \eqref{gc} has appeared before
 and many papers have been devoted to proving it in a variety of contexts.
It first appeared in the context of smooth dynamical systems where it concerned the fluctuations
of the phase space contraction, \cite{gc,ems,ru},
and the result has become known as the (steady state) fluctuation theorem.\\
We now prove \eqref{1c}.

\begin{proof}
By definition and since $J_i(\Theta\omega) = - J_i(\omega)$,
\[
\langle e^{-\lambda J_{i}}\rangle_\rho =  \int \id\bsP_{\rho}(\Theta
\omega)\, e^{\lambda J_{i}(\omega)}
\]
Next we insert the Radon-Nikodym derivative
\begin{equation}%
\int \id\bsP_{\rho}(\omega)\frac{\id\bsP_{\rho}\Theta}
{\id\bsP_{\rho}} (\omega)\,e^{\lambda J_{i}(\omega)} =
 \int
\id\bsP_{\rho}(\omega)\exp[\Delta+\delta J_{i}(\omega)]\,e^{\lambda
J_{i}(\omega)}%
\end{equation}%
where, via \eqref{girs} or via \eqref{conc},%
\begin{equation}%
\Delta=\log \frac{\rho (\omega _\tau)}{\rho (\omega _0)} + \beta \left(
H(\omega _{\tau})-H(\omega _{0})\right) -a\Delta {\cal N}+ \delta (J_{\ell}
(\omega)+J_{i} (\omega))%
\end{equation}%
Notice that $|\Delta {\cal N}| =|{\cal N}(\omega _\tau)-{\cal N}(\omega0)|\leq 2N$. Further,
$|\log \rho (\omega _\tau)/\rho(\omega_0)|$ is also bounded  because at any rate,
$\rho (\eta)\neq 0$, $\eta \in K$. Finally, there is the
conservation law
\begin{equation}%
J_{\ell} (\omega)+J_{i}
(\omega)=-N_{[-N,i]}(\omega _\tau)+N_{[-N,i]}(\omega _0)%
\end{equation}%
implying that the sum of these currents is also bounded, and
$|H(\omega _\tau)-H(\omega _0)|\leq 4(\kappa+B)N$, see
(\ref{hamilt}). Hence we conclude that $|\Delta|\leq \text{const}$ (with a constant that also
depends on
$N$ but not on
$\tau$) which finishes the proof.
\end{proof}

In the case $\beta=0$ (the bulk dynamics is that of the so called simple symmetric exclusion
process), more is known about the fluctuations of the current, see e.g.~\cite{der}.

\section{More nonequilibrium issues}

Only a limited review has been given in the previous sections of recent work on nonequilibrium
aspects of stochastic lattice gases. We attempt to give some additional remarks.
\subsection{Escape from equilibrium}
Imagine yourself enclosed in a well-isolated room.  At time zero somebody opens doors and windows;
how long would it take before you feel that?  Probably you will become aware of the openings
because of some air current, and its intensity will depend on the
outdoor conditions.\\
In the present section we use our model to ask a similar question. Suppose a probability
distribution $\mu$ on $K$ which is nonzero, $\mu(\eta) > 0$, only when ${\cal N}(\eta) = m$ for
some given number $0<m<N$ of particles.  For the rest we assume that it is thermally
distributed:\footnote{$\chi$ is the indicator function.}
\begin{equation}\label{cg}
\mu(\eta) = \frac 1{\caZ_m}\,\exp[-\beta H(\eta)]\;\chi(\cal N(\eta)
=m)
\end{equation}
We consider the dynamics of Section \ref{currents} in the steady state $\bsP_\rho$ and we ask for
the fraction of times that we see a fixed configuration
$\eta$:
\[
 p_\tau(\eta) = \frac 1{\tau}\,\int_0^\tau \chi(\eta_t=\eta)\,\id t
\]
That time-average is a random variable (depends on
$\eta$). An important question in the theory of large deviations is to ask whether these fractions
resemble a given probability measure; here we ask what is the function $I(\mu)$ so that
  \begin{equation}\label{ff}
\bsP_\rho[p_\tau \simeq \mu] \simeq e^{-\tau\, I(\mu)}, \qquad
\tau \uparrow +\infty
\end{equation}
That question has been rigorously studied by Donsker and Varadhan,
\cite{DZ,DV}, and we know an expression for $I(\mu)$:
\begin{equation}\label{dv}
I(\mu) = -\inf_{g>0}\Bigl\langle \frac{Lg}{g} \Bigr\rangle_\mu
\end{equation}
The expectation is over the distribution $\mu$ and $Lg$ is as before the generator of the process
acting on a function $g$ (over which we vary in \eqref{dv}).  It is known that in the case of a
detailed balance
 process, the minimizer in \eqref{dv} is $g^\star = \sqrt{\mu/\rho}$.\\

> From \eqref{dv} it is clear that we must find the function $g$ on
$K$ that minimizes
\[
\sum_{\eta\in K}\frac{\mu(\eta)}{g(\eta)} \Bigl[\sum_{i=1}^{N-1}
C(i,i+1,\eta)\, g(\eta^{i,i+1}) + g(\eta^{-N})\,C(-N,\eta) + g(\eta^N)\, C(N,\eta)\Bigr]
\]
The sum is effectively over all $\eta$ with $\mu(\eta) > 0$ or
$\cal N(\eta) = m$.  Since the configurations $\eta^{\pm N}$ have
one particle more or less than $\eta$, we can put $g(\xi)=0$ whenever $\cal N(\xi) \neq m$, and
the minimization is over
\[
\sum_{\eta\in K}\frac{\mu(\eta)}{g(\eta)} \sum_{i=1}^{N-1}
C(i,i+1,\eta)\, g(\eta^{i,i+1})
\]
Taking again the complete Donsker-Varadhan functional we have
\begin{equation}\label{im}
I(\mu) = \langle C(-N,\eta(-N)) + C(N,\eta(N))\rangle_\mu -
\inf_g \Bigl\langle \frac {\caL g}{g} \Bigr\rangle_\mu
\end{equation}
where the new generator $\cal L$ only considers particle exchanges generating a dynamics that
satisfies the condition of detailed balance with respect to our  $\mu$; in other words, it is
typical to `see' the distribution $\mu$ for that pure hopping process.  It follows that the second
term in \eqref{im} is zero to conclude that
\[
I(\mu)= \langle C(-N,\eta(-N)) + C(N,\eta(N))\rangle_\mu
\]

The issue here is somewhat related to the problem of metastability, as found also in the
contributions by Anton Bovier and by Frank den Hollander, see also \cite{BovierDenH}.  At the same
time, it is related to e.g.~Section 5.7 in \cite{DZ} (diffusion from a domain), and from a physics
point of view it is related to Kramers' theory, \cite{Hang}.

\subsection{Macroscopic fluctuations}

The previous calculation is related to the theory of dynamical fluctuations. There is however
another scale of description on which similar questions become better manageable, that is the
level of macroscopic fluctuations, static and dynamical.\\
For our model, it refers to a hydrodynamical scaling in which one observes the evolution of the
density profile.  After a diffusive space-time rescaling one finds that the density profile
$n_t(r)$ obeys a standard diffusion equation, \cite{kipnislandim},
\[
\frac{\partial n_t(r)}{\partial t} =\frac 1{2}
\frac{\partial}{\partial r} D(n_t(r)) \frac{\partial}{\partial
r}n_t(r)\,,\qquad r\in [-1,1]
\]
where $D(n_t(r))$ is the diffusion `constant,' further constrained by imposing the boundary
conditions $\rho(\pm 1) = 1/(1+e^{a_{\pm}})$.  In the simplest case (corresponding to
$\beta=0$) the diffusion is truly constant and the stationary
profile $n^\star$ is linear. One can however ask for fluctuations around that (typical) behavior.
The hydrodynamic equation is the result of a law of large numbers and we can ask for the
plausibility of a deviating density profile.  We refer to \cite{jona,der,KKM,brussel}
for further results \nolinebreak[4] and insights. \\

\newpage
\vspace{15mm}
\begin{center}
{\large{\bf Part II.\\ Macroscopic irreversibility}}
\end{center}
\vspace{3mm}

\addtocontents{toc}{\textbf{Part II. Macroscopic irreversibility}}

\section{Introduction}\label{sec: IntroII}

Up to now we have been discussing so called mesoscopic systems, or more precisely, classical
mesoscopic systems modeled as stochastic processes. Time-reversal symmetry was broken by applying
external conditions, frustrating the system in its return to equilibrium.  However, the
microscopic laws of nature are time-reversal invariant.  One could then perhaps have expected to
find that all resulting behavior is invariant under time-reversal, except perhaps for some
microscopic interactions.\footnote{All would probably agree e.g.~that the weak fundamental
interaction is not at all responsible for macroscopic irreversibility, and most would probably
agree that quantum mechanics is not either (while this is somewhat more tricky).} That is not what
we see: systems return to equilibrium thereby showing the infamous arrow of time. The equations of
macroscopic physics are not time-reversible (or not always).  They have often been described and
been used quite some time before their microscopic origin was clarified.  In fact their
(macroscopic) irreversibility once casted doubt on the kinetic and atomistic picture of matter and
motion.  One of the greatest successes in the  pioneering days  of statistical mechanics was then
indeed the explanation of
that manifest irreversibility.\\

That the emergent macroscopic laws are irreversible is not so difficult to understand at least
qualitatively.  One should realize that distinct macroscopic states can be very different in the
number of microstates they consist of. It is the installation of an initial macrostate that breaks
the invariance under time-reversal: unless forbidden by additional constraints, a less plausible
initial state evolves to a more plausible macrostate and finally to the most plausible, called
equilibrium,  exactly because that is more plausible. The plausibility is measured in terms of the
`number' of microstates or, more precisely, by the Boltzmann or counting entropy which has a well
defined thermodynamic limit (a precise meaning of that counting needs to be and will be
specified). The generic increase of the entropy between initial and final macrostates
(traditionally both in equilibrium) is known as the second law of thermodynamics and can
be formulated in various ways.\\
 Still, even more is often true:
considerations of entropy via counting the microstates consistent with a given macrostate, are
\emph{a priori} not restricted only to the initial and the final states and can be applied to each
intermediate, `nonequilibrium' state as well. Extended in that way, the Boltzmann entropy is often
an increasing function of time, as was first demonstrated for the Boltzmann equation, the
macroscopic evolution equation for rare gases, and rigorously proven in the so called
Boltzmann-Grad scaling limit and for short times by Lanford, \cite{Lanford}. Such a much more
detailed or `microscopic' version of the second law proves to be valid much beyond the Boltzmann
equation; for general
 theoretical arguments
  see~\cite{jay1,jay2,jay3,GL1,GL2,GM,prigogine}.\\

> From a mathematical point of view, the second law in the form of
an H-theorem in fact claims the existence of a Lyapunov functional for a class of evolution
equations, and it even hints at how to find that: if we know the underlying microscopic dynamics
from which the evolution equation (presumably or provably) follows, one is to search for the
Boltzmann entropy. Understanding why this strategy often works brings to the foreground some other
important observations: the validity of a macroscopic evolution equation means that there is a
\emph{typical} macroscopic behavior in the sense that it is a result of some law of large numbers.
The fact that the macroscopic equation is often first order in time means that this macroscopic
behavior is \emph{autonomous}.  On the other hand, the existence of microscopic configurations
violating that typical macroscopic law is not only allowed but in a sense it is even
\emph{necessary} for a true irreversible behavior and a strict increase of entropy to occur! When
formulated somewhat more precisely, these observations answer various apparent paradoxes as
formulated by Loschmidt and Zermelo; a qualitative discussion can be found on various places, see
e.g.\ \cite{Le,Br,jay1}.

Putting these arguments on a mathematically more precise level is relatively simple but it remains
very instructive. First, in Section~\ref{sec: Kac}, we study a model introduced by Mark Kac,
\cite{K}. The arguments are formulated in a substantially more
generality in Section~\ref{sec: generalizations} (the case of infinite dynamical systems) and
Section~\ref{sec: finite} (the case of large but finite systems). In particular, we explain how an
H-theorem follows from the very existence of an autonomous macroscopic dynamics. The resolution of
Zermelo's and Loschmidt's paradoxes is understood via a fluctuation symmetry, in this way drawing
a link to the first part of these lectures. Most of the presented material and some more details
can be found in~\cite{DMN1} and references therein.\\

Since the fundamental laws of nature are presumably quantum, one can further ask how the classical
arguments leading to H-theorems need to be changed
 when starting from a quantum microscopic dynamics.
There is no crucial difference up to one important point, namely that the very notion of a
macroscopic state needs to be reconsidered because it can be (and in nonequilibrium practice it
often \emph{is}) specified through values of mutually incompatible observables.  The
non-commutativity is a genuine quantum-mechanical feature which cannot be simply waived away by
arguments identifying the classical limit with the thermodynamic limit.  Furthermore, entropic
arguments and microscopic derivations all together play on the level of fluctuations, i.e.\ before
the thermodynamic limit. Starting from Section~\ref{sec: quantum}, we explain a possible approach
to the quantum problem along the lines of reference~\cite{DMN2}, thereby generalizing some ideas
of John von Neumann, \cite{vN}. An interesting side problem is to show how our construction of the
quantum Boltzmann entropy relates to other, mathematically simpler but physically
\emph{a priori} less plausible constructions. In fact, we show
when two definitions of quantum entropies become equivalent in the large system limit. That issue
is obviously very related to the problem of quantum large deviations and we will briefly describe
the connections. Finally, as an example, we come back to the Kac ring model and we discuss its
quantum extension along the lines of reference~\cite{qKac}; see Section~\ref{sec: qKac}.

\section{Kac ring model}\label{sec: Kac}

There is a simple paradigmatic model introduced by Mark Kac~\cite{K} to simplify the mathematics
of the Boltzmann equation. While the Boltzmann equation is much more complicated, the Kac model is
mathematically simple and free of those extra technical problems that are not really important for
understanding some crucial aspects of the emergence of macroscopic irreversibility.

\subsection{Microscopic dynamics}

Consider the set $\La = \{1,\ldots,N\}$. We imagine it as a ring in which we identify the sites
$1=N+1$ and on each site we have one particle and one scatterer.  The particles carry a `spin'
$\eta(i) =\pm 1$  and the scatterers can be off or on, $g(i) \in
\{0,1\}$. The resulting set $K = \{-1,1\}^\La \times \{0,1\}^\La$
is the state space of our model. The dynamics is deterministic and given via the transformation
$U$ on $K$,
\begin{equation}\label{eq: Kac-micro}
  (U(\eta,g))(i) = ([1 - 2g(i-1)]\,\eta(i-1),\, g(i)) \mod N
\end{equation}
which generates the (microscopic) deterministic dynamics such that the configuration
$(\eta_t,g_t)$ at time $t = 0,1,\ldots$ is
\begin{equation}
  \eta_t(i) = \eta_0(i-t) [1 - 2 g(i-t)]\ldots [1 - 2 g(i-1)]
\end{equation}
and $g_t = g$ keeps constant. There is an obvious interpretation: at every time instance $t$, each
spin
$\eta_t(i)$ jumps to its successive site, $i+1$, either
 flipping its value if a scatterer is
present, $g(i) = 1$, or keeping its value if $g(i) = 0$.
 Sampling the initial configuration
$(\eta_0,g)$ from a measure $\mu_0$ on $K$, the probability to
find $(\eta,g)$ at time $t$ is
\begin{equation}
  \mu_t[(\eta,g)] = \mu_0[(\eta_t,g)]
\end{equation}
That is the present variant of the Liouville equation for mechanical systems.  Here also the
Shannon entropy $S(\mu) = -\sum_{\eta,g} \mu[(\eta,g)] \log \mu[(\eta,g)]$ is time-invariant,
$S(\mu_t) = S(\mu_0)$. There is just no strictly increasing Lyapunov function for this dynamical
system;
 in fact,
 the dynamics is $2 N-$periodic. Nevertheless the
model exhibits relaxation to equilibrium; to see that, we need to pass to a macroscopic viewpoint.

\subsection{Macroscopic evolution}\label{eq: Kac-macroevolution}

There are two natural macroscopic observables, the magnetization
$m^N$ and the fraction of on-scatterers $\rho^N$:
\begin{equation}\label{eq: macroobservables}
  m^N = \frac{1}{N} \sum_{i=1}^N \eta(i)\,, \qquad
  \rho^N = \frac{1}{N} \sum_{i=1}^N g(i)
\end{equation}
The emergent \emph{macroscopic} dynamics will have the form
\begin{equation}
  (m^N_t,\rho^N) \mapsto (m^N_{t+1},\rho^N) = \phi (m_t^N,\rho^N)
\end{equation}
at least for very large $N$.  It would imply that the macroscopic data $(m^N_t,\rho^N_t)$ evolve
\emph{autonomously}, irrespectively of any actual microscopic configuration $(\eta_t,g_t)$ that
realize~\eqref{eq: macroobservables}. A simple heuristics\footnote{ Think of $N(1 \pm m)$ up
(down) spins crossing `on average' $N\rho$ scatterers every time step, entirely neglecting
possible time correlations. Such a hand-waving derivation is often referred to as
\emph{Stosszahlansatz} (or repeated randomization, molecular chaos approximation,...).} suggests
$\phi (m,\rho) = ([1 - 2\rho] m, \rho)$ as a candidate map. Yet, it is easy to imagine microscopic
configurations that violate that and the question arises how such a macroscopic
behavior can/must be understood.\\

Introducing the counting probability measures
\begin{equation}\label{eq: counting}
  \bbP^N[(\eta,g)] = 2^{-2N}
\end{equation}
and the notation $a \stackrel{\de}{=} b$ for $|a - b| \leq \de$, the desired statement has the
form of a law of large numbers:\footnote{ In fact, a strong law of large numbers is also true for
this model, see~\cite{K}, but the weak law is sufficient for our purposes. Later we will meet an
even substantially weaker autonomy condition.}
\begin{equation}\label{eq: Kac-autonomy}
  \lim_{N\uparrow\infty}
  \bbP^N[m^N(\eta_t)
   \stackrel{\de}{=} m_0 (1 - 2\rho)^t \rel m^N(\eta_0) =
   m_0;\, \rho^N(g) = \rho]
  = 1
\end{equation}
for all $\de > 0$. This means there is a set of
\emph{typical} microscopic configurations
satisfying the macroscopic law with map $\phi$; those configurations violating that law make a set
of zero limit measure. Such a macroevolution is called
 \emph{autonomous}; note that~\eqref{eq:
Kac-autonomy} is equivalent to
\begin{equation}\label{eq: Kac-autonomy1}
  \lim_{N\uparrow\infty}
  \bbP^N[\forall t \leq T:\,
  m^N(\eta_t) \stackrel{\de}{=} m_0 (1 - 2\rho)^t \rel m^N(\eta_0) = m_0;\, \rho^N(g) = \rho]
  = 1
\end{equation}
for all $\de > 0$ and any finite $T$.\\

The relaxation to equilibrium along that typical macroevolution is obvious by inspection but one
can also construct an explicit witness which is the
 \emph{Boltzmann entropy}
$s(m,\rho)$ defined as the large deviation rate function for the sequence
$(m^N, \rho^N)_{N\uparrow +\infty}$ of observables:
\begin{equation}
  \bbP^N[(m^N(\eta),\rho^N(g))] \simeq (m,\rho)] \simeq e^{N s(m,\rho)}
\end{equation}
This is to be understood in the logarithmic sense after taking the limit $N \uparrow +\infty$,
i.e., it is a shorthand for the limit statement
\begin{equation}
  s(m,\rho) = \lim_{\de\downarrow 0} \lim_{N\uparrow +\infty} \frac{1}{N} \log
  \bbP^N[m^N(\eta) \stackrel{\de}{=} m;\,\rho^N(\eta) \stackrel{\de}{=} \rho]
\end{equation}
This is simply the binomial entropy,
\begin{equation}\label{eq: Kac-entropy}
  s(m,\rho) =
\begin{cases}
  - \frac{1+m}{2} \log (1+m) - \frac{1-m}{2} \log (1-m)\\
  \hspace{8mm} -\rho \log 2\rho - (1-\rho) \log 2(1-\rho)
  & \text{if } -1 < m < 1,\, 0 < \rho < 1
\\
  -\infty & \text{otherwise}
\end{cases}
\end{equation}
and one checks that $s(\phi(m,\rho)) > s(m,\rho)$ whenever $m \neq 0$ (system off equilibrium) and
$0 < \rho < 1$ (nonsingular macrodynamics). Using the notation $m_t = m_0 (1-2\rho)^t$, it yields
that $s(m_t,\rho)$ is a strictly increasing function
 of time. Following Boltzmann's
terminology, such a  statement is called
 an \emph{H-theorem}; the Boltzmann entropy is a Lyapunov function.

It should be clear that we say nothing yet about possible macroevolutions corresponding to those
exceptional microstates not verifying the macroscopic map
$\phi$. That will come in Section~\ref{sec: irreversibility}.\\

\emph{Proof~\eqref{eq: Kac-autonomy}}: Use the shorthand
\[
  \tilde\bbE^N[\,\cdot\,] = \bbE^N[\,\cdot \rel m^N(\eta_0) = m_0;\, \rho^N(g) = \rho]
\]
for the expectations conditioned on the initial macrostate
$(m_0,\rho)$. One easily checks that (1) $\bbP^N$ are
permutation-invariant measures, (2) $\eta_0$ and $g$ are independently distributed under $\bbP^N$,
and (3) the following asymptotic decoupling is true:\footnote{ This property can be recognized as
an instance of the equivalence between microcanonical and canonical ensembles.} provided that $-1
< m_0 < 1$ and $0 < \rho < 1$, there is a sequence $\De_N^{(k)}$, $\lim_N
\De_N^{(k)} = 0$ for all $k = 1,2,\ldots$, such that
\begin{align}\label{eq: Kac-bound1}
  |\tilde\bbE^N[\eta(i_1)\ldots\eta(i_k)]
  - (\tilde\bbE^N[\eta(1)])^k| &\leq \De_N^{(k)}
\\ \intertext{and}\label{eq: Kac-bound2}
  |\tilde\bbE^N[g(i_1)\ldots g(i_k)]
  - (\tilde\bbE^N[g(1)])^k| &\leq \De_N^{(k)}
\end{align}
for all $1 \leq i_1 < i_2 < \ldots < i_k \leq N$

Then, one subsequently gets
\begin{equation}\label{eq: moment1}
\begin{split}
  \tilde\bbE^N[m^N(\eta_t)] &=
  \frac{1}{N} \sum_{i=1}^N \tilde\bbE^N[\eta_0(i-t) \{1-2g(i-t)\}\ldots\{1-2g(i-1)\}]
\\
  &= \tilde\bbE^N[\eta_0(1)]\,\tilde\bbE^N[\{1-2g(1)\}\ldots\{1-2g(t)\}]
\\
  &= m_0 \{(1-2\,\tilde\bbE^N[g(1)])^t + A_N\}
\\
  &= m_0(1-2\rho)^t + o(1)
\end{split}
\end{equation}
by using that $|A_N| \leq \De_N^{(t)}$ due to~\eqref{eq: Kac-bound2}. Similarly for the second
moment,
\begin{equation}
\begin{split}
  \tilde\bbE^N[(m^N(\eta_t))^2] &=
  \frac{1}{N} \sum_{i=1}^N \tilde\bbE^N[\eta_0(1)\, \eta_0(i)]\,
  \tilde\bbE^N[\{1-2g(1)\}\ldots\{1-2g(t)\}
\\
  &\hspace{5mm} \times \{1-2g(i)\}\ldots\{1-2g(i+t-1)\}]
\end{split}
\end{equation}
Observe that for $t+1 \leq i \leq N+1-t$ there is no pair of the $g$'s in the above product,
acting on the same site. Hence, using the permutation-invariance and the asymptotic
decoupling~\eqref{eq: Kac-bound1}--\eqref{eq: Kac-bound2},
\begin{equation}\label{eq: moment2}
\begin{split}
  \tilde\bbE^N[(m^N(\eta_t))^2] &=
  \frac{1}{N} \sum_{i=t+1}^{N+1-t} \{m_0^2 + B_N(i)\}
  \{(1 - 2\,\tilde\bbE^N[g(1)])^{2t} + C_N(i)\}
\\
  &\hspace{5mm}+ \frac{1}{N} \Bigl(\sum_{i=1}^t + \sum_{i=N+2-t}^N \Bigr)
  \{m_0^2 + B_N(i)\}
\\
  &\hspace{5mm}\times\tilde\bbE^N[\{1-2g(1)\}\ldots\{1-2g(t)\}
\\
  &\hspace{10mm} \times \{1-2g(i)\}\ldots\{1-2g(i+t-1)\}]
\\
  &= \frac{N-2t+1}{N}m_0^2 (1-2\rho)^{2t} + o(1)
\\
  &= m_0^2 (1-2\rho)^{2t} + o(1)
\end{split}
\end{equation}
since the remainders satisfy $|B_N(i)| \leq \De_N^{(2t)}$,
$|C_N(i)| \leq \De_N^{(2t)}$ for all $i$, and using a simple bound on the last term. The weak law
of large numbers~\eqref{eq: Kac-autonomy} then follows from~\eqref{eq: moment1} and
\eqref{eq: moment2} via a Chebyshev inequality.\\

Remark that there is a considerable freedom in the choice of the measures from which the initial
configurations are sampled. The `microcanonical' measure
\begin{equation}
  \tilde\bbP^N[\,\cdot\,] = \bbP^N[\,\cdot \rel m^N(\eta) = m_0;\,\rho^N(g) = \rho]
\end{equation}
is most natural but for obtaining~\eqref{eq: Kac-autonomy}, it can be replaced by various other
ensembles.  There is for example the `canonical' measure
\begin{equation}
  \bbP^N_\can[(\eta,g)] =
  \frac{1}{\caZ^N}\exp \sum_{i=1}^N (\be \eta(i) + \al g(i))
\end{equation}
with the Lagrange multipliers $\be,\al$ being fixed by the conditions
\[
  \bbE^N_\can[\eta(1)] = m_0\,,\qquad
  \bbE^N_\can[g(1)] = \rho
\]
and $\caZ^N$ is the normalization factor.
 It is easy to check that the autonomy~\eqref{eq:
Kac-autonomy} remains true if replacing
$\tilde\bbP^N$ with $\bbP^N_\can$; in fact,
the proof is simpler now since $\bbP^N_\can$ exactly factorizes and hence the above remainders
$A_N$, $B_N$, $C_N$ as in~\eqref{eq: moment1} and
\eqref{eq: moment2}
are zero.\\
The corresponding large deviation rate function $s_\can(\eta,g)$ which enters the law
\begin{equation}
  \bbP^N_\can[(m^N(\eta),\rho^N(g)) \simeq (m,\rho)] \simeq e^{N s_\can(m,\rho)}
\end{equation}
can also be computed (the easiest via the Gartner-Ellis theorem, \cite{DZ}) with the result
$s_\can(m,\rho) = s(m,\rho)$. This equality shows
 the equivalence of ensembles on the level of
entropies, which is well studied in equilibrium statistical physics. We come back to the problem
of equivalence within a quantum framework and in a substantially larger generality in
Section~\ref{sec: equivalence}.

\subsection{Irreversibility and entropy production}\label{sec: irreversibility}

Consider a modification of the microdynamics~\eqref{eq: Kac-micro} in which the particles jump to
the left instead of to the right. It is given by the map
\begin{equation}
  (\bar U(\eta,g))(i) = ([1 - 2g(i)]\,\eta(i+1),\, g(i)) \mod N
\end{equation}
which is an inverse of $U$, i.e., $\bar U \circ U = U \circ \bar U = 1$. The spin configuration at
time $t$ as evolved from $\eta$ through the dynamics $\bar U$ is denoted by
$\bar\eta_t$:
\begin{equation}
  \bar\eta_t(i) = \eta_0(i+t) [1 - 2 g(i+t-1)]\ldots [1 - 2 g(i)]
\end{equation}
Observe that the sequence (trajectory)
$(\eta_0,\eta_1,\ldots,\eta_t)$ is allowed (possible) under the
original microscopic dynamics iff
$(\eta_t,\eta_{t-1},\ldots,\eta_0)$ is possible under $U$.  That
invertibility is referred to as \emph{dynamical (time-)reversibility}.  It can be formulated
differently by extending the configuration space $K$ with a `velocity' variable
$v \in \{-1,1\}$ and by defining the dynamics via the
transformation
\begin{equation}
  V(\eta,g,v) =
\begin{cases}
  (U(\eta,g),v) & \text{if } v = +1
\\
  (\bar U(\eta,g),v) & \text{if } v = -1
\end{cases}
\end{equation}
With the involution
 $\pi$, $\pi(\eta,g,v) = (\eta,g,-v)$ the dynamical
  reversibility gets the
form: $\pi \circ V \circ \pi = V^{-1}$; the
 time-reversed microscopic dynamics is then achieved by
inverting the velocity $v$.\\

The macroscopic time-evolution $\phi$ in which $m \mapsto (1-2\rho)m \in [-1,1]$ is invertible as
well (provided that $\rho \neq \frac{1}{2}$). Yet, the typical macroscopic time-evolution
corresponding to
$\bar U$ is not $\phi^{-1}$ but rather $\phi$ again, i.e., the law of large numbers~\eqref{eq: Kac-autonomy}
stays true when replacing $U$ and
$\eta_t$ with $\bar U$ and $\bar\eta_t$! It simply means that the
macroscopic evolution $m \mapsto m (1 - 2\rho)$ does \emph{not} get inverted by starting from a
\emph{typical} microscopic configuration $\eta$ corresponding to the macroscopic state $m
(1-2\rho)$ and by applying the inverted microscopic dynamics (or by inverting the velocity).
Naturally, there exist microscopic configurations $\eta$ for which the inverted macroevolution $m
(1-2\rho) \mapsto m$ along the dynamics $\bar U$ would be observed---this is precisely what the
dynamical reversibility claims---they show up to be exceedingly
\emph{exceptional} under $m (1-2\rho)$, however. This physical
impossibility to invert the macroscopic evolution is referred to as
\emph{macroscopic irreversibility}.\\

The macroscopic irreversibility in the above sense on the one hand and the strict increase of the
Boltzmann entropy on the other hand, are often used as synonyms. Let us formulate their relation a
bit more precisely. Observe that the two sets
\[
  \{(\eta_0,g):\,m^N(\eta_0) \stackrel{\de}{=} m_0;\,m^N(\eta_t)
  \stackrel{\de}{=} m_t;\,\rho^N(g) \stackrel{\de}{=} \rho\}
\]
and
\[
  \{(\eta_0,g):\,m^N(\bar\eta_t) \stackrel{\de}{=} m_0;\,m^N(\eta_0)
  \stackrel{\de}{=} m_t;\,\rho^N(g) \stackrel{\de}{=} \rho\}
\]
with $m_t = m_0 (1-2\rho)^t$ have the same cardinalities (check that the map $\bar U$ is a
bijection between these sets). Hence, they have the same measures under $\bbP^N$, which we write
as
\begin{equation}\label{eq: inverted1}
\begin{split}
  \log \bbP^N[m^N(\eta_t) \stackrel{\de}{=} m_t \rel m^N(\eta_0)
  &\stackrel{\de}{=} m_0;\,\rho^N(g) \stackrel{\de}{=} \rho]
\\
  +\log \bbP^N[m^N(\eta_0) &\stackrel{\de}{=} m_0;\,\rho^N(g)
  \stackrel{\de}{=} \rho]
\\
  = \log \bbP^N[m^N(\bar\eta_t) \stackrel{\de}{=} m_0 \rel m^N(\eta_0)
  &\stackrel{\de}{=} m_t;\,\rho^N(g) \stackrel{\de}{=} \rho]
\\
  + \log \bbP^N[m^N(\eta_0) &\stackrel{\de}{=} m_t;\,\rho^N(g)
  \stackrel{\de}{=} \rho]
\end{split}
\end{equation}
Dividing by $N$, taking the limits $N\uparrow\infty$ and $\de\downarrow 0$ in this order, and
using the law of large numbers~\eqref{eq: Kac-autonomy}, we get the large deviation law
\begin{equation}\label{eq: inverted2}
\begin{split}
  \lim_{\de\downarrow 0} &\lim_{N\uparrow\infty}
  \frac{1}{N} \log\bbP^N[m^N(\bar\eta_t) \stackrel{\de}{=} m_0
  \rel m^N(\eta_0) \stackrel{\de}{=} m_t;\, \rho^N(g) \stackrel{\de}{=} \rho]
\\
  &= s(m_0,\rho) - s(m_t,\rho)
\end{split}
\end{equation}
or
\begin{equation}\label{eq: main-macro}
  \bbP^N[m^N(\bar\eta_t) \simeq m_0 \rel m^N(\eta_0) \simeq m_t;\,\rho^N(g) \simeq \rho]
  \simeq e^{-N[s(m_t,\rho) - s(m_0,\rho)]}
\end{equation}
which is quite a remarkable relation. Notice first that it provides another derivation of the
H-theorem: since the left-hand side is less than one, one immediately gets $s(m_t,\rho) \geq
s(m_0,\rho)$. Further, the left-hand side is nothing but the probability that a configuration
$\eta_0$ sampled from macrostate
$m_t$ and evolved according to $\bar U$, exhibits the macroscopic
transition from $m_t$ to $m_0$, which is just an inversion of the typical transition $m_0 \mapsto
m_t$.

Macroscopic irreversibility amounts to the statement that such inverted macroscopic transitions
are physically impossible; here we have a quantitative evaluation how rare they really are: the
large deviation rate function for the backward transition $m_t
\mapsto m_0$ with respect to $\bar U$ just coincides with the
entropy production along the \emph{typical} macroevolution $m_0
\mapsto m_t$ with respect to microscopic dynamics $U$. Inverting
the logic, this can be read off as a formula for the entropy production, possibly useful provided
that
the probability of those rare backward transitions can be evaluated or estimated.\\

The seeming inconsistency between the microscopic reversibility and the macroscopic
irreversibility is known as the Loschmidt paradox. Equality~\eqref{eq: main-macro} in a sense
solves this paradox and put it in a correct perspective: those macroscopic trajectories obtained
by time-reverting the typical ones
$(\phi^t(m)$;\, $t = 0,1,\ldots)$ are indeed observable for finite
$N$, however they are exponentially damped. Notice that~\eqref{eq:
main-macro}  is nothing but a macroscopic analogue of the detailed balance
condition~\eqref{detailbalance} that we have already discussed in the context of lattice gases.

\section{Infinite systems}\label{sec: generalizations}

The above analysis of the Kac model shows up to be quite generic and one can easily extend those
arguments to a more general setup. The aim of the present section is to formulate general
sufficient conditions for
 the existence of a Lyapunov function for a class of macroscopic
dynamics, or, equivalently, for an H-theorem to be valid.

\subsection{Dynamical systems, macrostates, and entropy}\label{sec: dynamical systems}

On a microscopic level, we consider a family of classical dynamical systems
$(K^N,U^N_t,\bbP^N)_{N\uparrow +\infty}$, where the label $N$ should be thought of as a spatial
extension of the system and the maps\footnote{Depending on an application, the maps $U^N_t$ can be
e.g.\ Hamiltonian flows, possibly with the time and space suitably rescaled with $N$. The details
are not really important for what follows, we will only require the microscopic dynamics to
satisfy a few general conditions, see below.}
$(U^N_t)_{t \geq 0}$ are assumed to satisfy the semigroup condition
$U^N_t\, U^N_s = U_{t+s}$ for all $t,s \geq 0$. The probability measures $\bbP^N$
are invariant under the dynamics:
$\bbP^N (U^N)^{-1} = \bbP^N$.\\

The macroscopic level of description is specified by a collection of macroscopic observables.
These are some maps $M^N: K^N \mapsto \Om$ into a metric space $(\Om,d)$ of \emph{macrostates}. We
assign to every $m \in \Om$ the Boltzmann entropy defined as the large deviation rate function
under the measures
$\bbP^N$:
\begin{equation}\label{eq: macrospace}
  s(m) = \lim_{\de\downarrow 0} \limsup_{N\uparrow +\infty}
  \frac{1}{N} \log\bbP^N[M^N(x) \stackrel{\de}{=} m]\,, \qquad m \in \Om
\end{equation}
using the shorthand $m \stackrel{\de}{=} m'$ whenever
 $d(m,m') \leq \delta$. Denote
\begin{equation}
  \Om_0 = \{m \in\Om;\,s(m) > -\infty\}
\end{equation}
the set of those macrostates that are admissible; we assume $\Om_0 \neq \emptyset$. Note this is a
first nontrivial assumption: the parametrization by $N$ and an eventual rescaling have to be
meaningful so that $(M^N)$ indeed satisfies the large deviation principle with a finite rate
function $s(m)$ on some large enough space $\Om_0$.

\subsection{Autonomous evolution and H-theorem}\label{sec: H-general}

Starting from a microscopic configuration $x \in K^N$, the macroscopic trajectory is simply the
collection
$(M^N(U^N_t x))_{t\geq 0}$. We assume the existence of an \emph{autonomous}
macroscopic dynamics in the following sense: let there be a collection $(\phi_t)_{t\geq 0}$ of
maps $\phi: \Om_0 \mapsto \Om_0$ satisfying
\begin{enumerate}
\item
the semigroup condition
\begin{equation}\label{eq: semigroup}
  \phi_t \circ \phi_s = \phi_{t+s}\,,\qquad t,s \geq 0
\end{equation}
\item a weak autonomy condition
\begin{equation}\label{eq: autonomy}
  \lim_{\de\downarrow 0} \lim_{N\uparrow +\infty} \frac{1}{N}
  \log\bbP^N[M^N(U^N_t x) \stackrel{\de}{=} \phi_t(m) \rel
  M^N(x) \stackrel{\de}{=} m] = 0
\end{equation}
for all $t \geq 0$ and $m \in \Om_0$.
\end{enumerate}
Notice that~\eqref{eq: autonomy} is a much weaker condition than the law of large
numbers~\eqref{eq: Kac-autonomy} valid for the Kac model. In particular, no typical macroscopic
evolution is required to exist; that $(\phi_t)_{t\geq 0}$ can e.g.\ be a single realization of a
stochastic process describing a macroscopic evolution of a system passing through a number of
branching points.\footnote{To have in mind a specific scenario, think of a ferromagnet being
cooled down from a high-temperature paramagnetic state. When passing the critical temperature, the
system randomly (= depending on the initial microscopic configuration) chooses one of the
ferromagnetic states with broken symmetry.} On the other hand, this condition is generally not
satisfied by stochastic systems on mesoscopic scale and/or without involving
 the large $N$ limit. In that
sense, condition~\eqref{eq: autonomy} draws a sharp border line
between macroscopic and mesoscopic systems.\\

Since $\bbP^N$ is invariant under $U^N_t$, conditions~\eqref{eq: semigroup}--\eqref{eq: autonomy}
are equivalent with a single condition
\begin{equation}\label{eq: autonomy1}
  \lim_{\de\downarrow 0} \lim_{N\uparrow +\infty} \frac{1}{N}
  \log\bbP^N[M^N(U^N_t x) \stackrel{\de}{=} \phi_t(m) \rel
  M^N(U_s x) \stackrel{\de}{=} \phi_s(m)] = 0
\end{equation}
required for all $t \geq s \geq 0$. Using the invariance $\bbP^N$ again, we find that for any pair
$m,m' \in \Om_0$ of macrostates,
\begin{equation}\label{eq: bounds}
\begin{split}
  \log{\bbP^N (M^N( x) \stackrel{\de}{=} m' )} &=
  \log{\bbP^N (M^N( U^N_t x) \stackrel{\de}{=} m' )}
\\
  &\geq \log{\bbP^N (M^N( U^N_t x) \stackrel{\de}{=} m'
  \rel M^N( U_s^N x) \stackrel{\de}{=} m )}
\\
  &\hspace{5mm}+ \log \bbP^N(M^N( U_s^N x) \stackrel{\de}{=} m)
\end{split}
\end{equation}
which we are again going to divide by $N$, to take the upper limit $N\uparrow +\infty$ and then to
take the limit $\de\downarrow 0$. Choosing first
$m = \phi_s(m)$ and $m' = \phi_t(m)$, autonomy condition~\eqref{eq: autonomy1} yields
\begin{equation}\label{eq: H-general}
  s(\phi_t(m)) \geq s(\phi_s(m))\,,\qquad t \geq s \geq 0
\end{equation}
which is an H-theorem. Second, for $m = \phi_t(m)$ and $m' = \phi_s(m)$ it yields the inequality
\begin{multline}\label{eq: ep-upperbound}
  -\lim_{\de\downarrow 0} \limsup_{N\uparrow +\infty}
  \frac{1}{N} \log{\bbP^N (M^N( U^N_t x) \stackrel{\de}{=} \phi_s(m)
  \rel M^N( U_s^N x) \stackrel{\de}{=} \phi_t(m) )}
\\
  \geq s(\phi_t(m)) - s(\phi_s(m))
\end{multline}
again for all $t \geq s \geq 0$, which provides an upper bound on the Boltzmann entropy
production.\\

For an invertible microdynamics $(U^N_t)$ inequalities~\eqref{eq: H-general}--\eqref{eq:
ep-upperbound} can be turned into a single equality by essentially repeating the computation of
Section~\ref{sec: irreversibility}, see~\eqref{eq: inverted1}--\eqref{eq: inverted2}. The result
reads
\begin{equation}\label{resrea}
  s(\phi_t(m)) - s(\phi_s(m)) = \bar J_{s,t}(\phi_t(m), \phi_s(m)) \geq 0
\end{equation}
where
\begin{multline}
  -\bar J_{s,t}(m,m')
\\
  = \lim_{\de\downarrow 0} \limsup_{N\uparrow +\infty}
  \frac{1}{N} \log \bbP^N[M^N(\bar U_t^N x) \stackrel{\de}{=} m' \rel
  M^N(\bar U_s^N x) \stackrel{\de}{=} m]
\end{multline}
is the rate function for the transition from macrostate $m$ at time $s$ to macrostate $m'$ at time
$t$ along the time-reversed dynamics $\bar U^N_t \equiv (U^N_t)^{-1}$. The conclusions of
Section~\ref{sec: irreversibility} apply as well in this general case:
$\bar J_{s,t}(\phi_t(m),\phi_s(m))$ is a natural `measure' of macroscopic irreversibility, and we
have proven it is just equal to the entropy production for the transition
$\phi_s(m) \mapsto \phi_t(m)$ fulfilling the autonomy~\eqref{eq: autonomy1}.\\

Semigroup condition~\eqref{eq: semigroup} is crucial and cannot be simply relaxed. Indeed,
assuming only the autonomy in the form~\eqref{eq: autonomy}, one would still have the inequality
between the initial and final Boltzmann entropies: $s(\phi_t(m)) \geq s(m)$, $t \geq 0$, however,
$s(m_t)$ might not be monotone in general. As an example, think of the macrodynamics
$\phi_t: \bbR \mapsto \bbR$ given as
$\phi_t(m) = m\, r^t \cos \om t$, $|r| < 1$, which is like the position of an underdamped pendulum
swinging around its equilibrium position.

The missing semigroup property can be recovered by including additional macroscopic observables;
in the case of the pendulum one would naturally add its velocity as another observable. To
conclude, the semigroup condition is basically a restriction on the choice of the collection of
macroscopic observables, which needs to be in that sense `complete.'


\section{Finite systems}\label{sec: finite}

In this section we evaluate the necessity of the large $N$ limit in the above arguments, and we
attempt a microscopic formulation of the
 H-theorem for an entropy defined upon a finite system
and as a functional on microscopic configurations.

\subsection{Zermelo-Poincar\'e paradox}\label{sec: finite-intro}

For any fixed $N$ the dynamical system
$(K,U_t,\bbP) \equiv (K^N,U^N_t,\bbP^N)$ is really a \emph{finite size} system; this is encoded by
the assumption that $\bbP$ is a probability (and hence normalizable) measure. For simplicity, we
consider in this section a macroscopic observable $M: K \mapsto \Om$ such that
\begin{equation}
  \Om_0 = \{m \in \Om;\,\bbP[M^{-1}(m)] > 0\}
\end{equation}
is finite or countable. As entropy function we take
\begin{equation}\label{eq: entropy-finite0}
  S(m) = \log\bbP[M^{-1}(m)]\,, \qquad m \in \Om_0
\end{equation}

The well-known Poincar\'e recurrence theorem then reads that for
$\bbP-$almost every microstate $x \in K$ and any time $t_0$ one has
$M(U_t x) = M(x)$ for some $t > t_0$, i.e., the trajectory almost
surely returns back to the initial macrostate $M(x)$. This is usually phrased as the impossibility
for the entropy~\eqref{eq: entropy-finite} to be an increasing (nonconstant) function of time,
which is also known as the Zermelo(-Poincar\'e) paradox. However, the analysis in the previous
section and the Kac example give a clue: the recurrence time increases with the system size
$N$ and is shifted away to infinity in the thermodynamic
limit.\footnote{For the Kac example the recurrence time is of order $N$. However, in many
dynamical systems it goes exponentially with the number of degrees of freedom, which
is itself $e^{O(N)}$.}\\

In the context of H-theorems it is instructive to reformulate Zermelo's objection in still a
slightly modified way. There is in fact a tempting `trivialization' of the argument leading to the
H-theorem which goes as follows:

For our finite system, the autonomy could simply mean that the macroscopic evolution as specified
by a map $\phi: \Om_0 \mapsto
\Om_0$ is just what takes place for almost every microstate, i.e.,
that for $\bbP-$a.e.\ $x \in K$ one has
\begin{equation}\label{eq: autonomy-bad}
 M(U_t x) = \phi_t(M(x))\,, \qquad t \geq 0
\end{equation}
Were this indeed true as such, it would automatically imply the semigroup condition since, almost
surely,
\begin{equation}
  M(U_{t+s} x) = \phi_t(M(U_s x)) = \phi_t \circ \phi_s(M(x))\,,\qquad
  t,s \geq 0
\end{equation}
Second, it would mean that
\begin{equation}\label{eq: autonomy-bad1}
  \bbP[(U_t)^{-1} M^{-1}(m_t) \cap M^{-1}(m)] = \bbP[M^{-1}(m)]
\end{equation}
with $m_t = \phi_t(m)$, i.e., the set of microstates $M^{-1}(m)$ evolves to a subset of
$M^{-1}(m_t)$, up to a zero measure set. Hence,
\begin{equation}\label{finmon}
  S(m_t) = \log \bbP[(U_t)^{-1} M^{-1}(m_t)] \geq \bbP[M^{-1}(m)] = S(m)
\end{equation}
due to the invariance of $\bbP$. Finally,
\begin{equation}
  S(m_t) = S(\phi_{t-s}(m_s)) \geq S(m_s)\,,\qquad t \geq s \geq 0
\end{equation}
On the other hand, by the above Poincar\'e recurrence, $S(m_t) = S(m)$ for infinitely many $t$,
and
hence $S(m_t)$ is constant!\\

The above computation shows that the assumption of autonomy in the form~\eqref{eq: autonomy-bad}
or \eqref{eq: autonomy-bad1} is too strong.  If fulfilled, the macroscopic evolution would
necessarily be reversible and the entropy constant. Our condition of autonomy~\eqref{eq: autonomy}
is much weaker and it does not guarantee the semigroup property.  In the Kac example, the law of
large number~\eqref{eq: Kac-autonomy} is far stronger than autonomy~\eqref{eq: autonomy}, yet
still consistent with a macroscopic irreversible evolution, as we have checked explicitly.

\subsection{Microscopic H-theorem}\label{sec: H-general}

We come back to the general framework of Section~\ref{sec: dynamical systems} and consider again a
sequence of dynamical systems $(K^N,U^N_t,\bbP^N)_N$ and a general macroscopic observable $M^N:
K^N \rightarrow \Om$, macroscopic in the sense that $\Om_0$ defined in~\eqref{eq: macrospace} is
nonempty. Our aim is now to formulate an H-theorem for Boltzmann entropy assigned to each
\emph{microstate} of a single, possibly large finite-size system. Put it differently, we want to
see how much the entropy is allowed to fluctuate around a monotone path when evaluated along a
single microscopic
trajectory of a single dynamical system with $N$ large but fixed.\\

For $N$ fixed the Boltzmann entropy is no longer unambiguously defined since the sets
$\{x;\,M^N(x) \stackrel{\de}{=} m\}$ depend on the width $\de > 0$ and are not necessarily all
disjoint for different macroscopic states $m$. To be safe we assign to every microstate $x \in
K^N$ the interval
$[S_<^{N,\de}(x),S_>^{N,\de}(x)]$ of entropies defined as
\begin{align}
  S_<^{N,\de}(x) &= \inf_{m \in \Om_0} \{S^{N,\de}(m);\, m
  \stackrel{\de}{=} M^N(x)\}
\\
  S_>^{N,\de}(x) &= \sup_{m \in \Om_0} \{S^{N,\de}(m);\,
   m \stackrel{\de}{=} M^N(x)\}
\end{align}
where
\begin{equation}\label{eq: entropy-finite}
  S^{N,\de}(m) = \log\bbP^N[M^N(x) \stackrel{\de}{=} m]
\end{equation}
We impose the autonomy assumption in the form of a law
 of large numbers:
\begin{equation}\label{eq: autonomy-strong}
  \lim_{\de\downarrow 0} \lim_{N\uparrow +\infty}
  \bbP^N[M^N(U^N_t x) \stackrel{\de}{=} \phi_t(m) \rel
  M^N(x) \stackrel{\de}{=} m] = 1
\end{equation}
for all $m \in \Om_0$, $t \geq 0$, and with maps $\phi_t:\Om_0 \mapsto \Om_0$ such that
$\phi_t \circ \phi_s = \phi_{t+s}$, $t \geq s \geq 0$.\\

Let us fix some initial condition $m \in \Om_0$, $\de > 0$ and a finite sequence of times $0 = t_0
< t_1 < \ldots < t_Q$. Combining assumption~\eqref{eq: autonomy-strong} with the semigroup
condition, the remainder
\begin{equation}\label{eq: reminder-def}
  D^{N,\de}(s,t;m) := 1 - \bbP^N[M^N(U^N_t x) \stackrel{\de}{=} \phi_t(m) \rel
  M^N(U^N_s x) \stackrel{\de}{=} \phi_s(m)]
\end{equation}
satisfies $\lim_{\de\downarrow 0} \lim_{N\uparrow +\infty} D^{N,\de}(s,t;m) = 0$ whenever
$t \geq s \geq 0$. By subadditivity,
\begin{multline}\label{eq: reminder}
  \bbP^N[M^N(U^N_{t_j} x) \stackrel{\de}{=}
  \phi_{t_j}(m),\, j = 1,\ldots,Q
  \rel M^N(x) \stackrel{\de}{=} m]
\\
  \geq 1 - \sum_{j=1}^Q D^{N,\de}(0,t_j;m)
\end{multline}
Whenever $m^N(U_{t} x) \stackrel{\de}{=} \phi_{t}(m)$ then
\begin{equation}
  S_<^{N,\de}(U^N_t x) \leq S^{N,\de}(\phi_t(m)) \leq S_>^{N,\de}(U^N_t x)
\end{equation}
As a consequence, \eqref{eq: reminder} gives
\begin{multline}\label{eq: finite-bound1}
  \bbP^N[S_<^{N,\de}(U_{t_j} x) \leq S^{N,\de}(\phi_{t_j}(m))
  \leq S_>^{N,\de}(U_{t_j} x),\, j=1,\ldots,Q
  \rel M^N(x) \stackrel{\de}{=} m]
\\
  \geq 1 - \sum_{j=1}^Q D^{N,\de}(0,t_j;m)
\end{multline}
Entropies at successive times satisfy the inequality following from~\eqref{eq: reminder-def}:
\begin{equation}\label{eq: finite-bound2}
  S^{N,\de}(\phi_{t_j}(m)) \geq S^{N,\de}(\phi_{t_{j-1}}(m))
  + \log(1 - D^{N,\de}(t_{j-1},t_j;m))
\end{equation}
Using that
\begin{equation}
  \lim_{\de\downarrow 0} \lim_{N\uparrow +\infty} \min_{j=1}^Q
  \log(1 - D^{N,\de}(t_{j-1},t_j;m)) = 0
\end{equation}
inequalities~\eqref{eq: finite-bound1}--\eqref{eq: finite-bound2} yield the main result of this
section:

For any $\De > 0$, $m \in \Om_0$, and a finite sequence $0 = t_0 < t_1 < \ldots t_Q$ of times, one
has
\begin{equation}\label{eq: finite-H}
  \lim_{\de\downarrow 0} \lim_{N\uparrow +\infty} \bbP^N
  [S_>^{N,\de}(U_{t_j}x) \geq S_<^{N,\de}(U_{t_{j-1}}x) - \De,\, j = 1,\ldots,Q
  \rel M^N(x) \stackrel{\de}{=} m] = 1
\end{equation}
Therefore, the finite-system entropy violates the monotonicity as little as required with
probability arbitrarily closed to one, provided that $\de$ is small enough and $N$ large enough.
This is the announced microscopic H-theorem.\\

Remark that the ambiguity with the definition of finite-system entropy does not arise if the
observables $M^N$ take only finitely or countably many values (as in Section~\ref{sec:
finite-intro} but this time uniformly for all $N$). In that case, one can set $\de = 0$ and the
entropy is simply
\begin{equation}
  S^N(x) := S_<^{N,0}(x) = S_>^{N,0}(x) = \log \bbP^N[(M^N)^{-1} M^N(x)]
\end{equation}
(Compare with \eqref{eq: entropy-finite0}.) The microscopic H-theorem~\eqref{eq: finite-H} then
becomes, under the same assumptions,
\begin{equation}
  \lim_{N\uparrow +\infty} \bbP^N
  [S^{N}(U_{t_j}x) \geq S^{N}(U_{t_{j-1}}x) - \De,\, j = 1,\ldots,Q
  \rel M^N(x) = m] = 1
\end{equation}

%

\section{Quantum systems}\label{sec: quantum}

The arguments presented in the previous section do not dramatically change when passing from a
classical
 to a quantum dynamics. What does need to be refined however, is the very
description of macroscopic states due to the inherent incompatibility of quantum observables, as
visible from the noncommutativity of corresponding (self-adjoint) operators before any macroscopic
limit is taken.

Obviously, the question of a quantum fluctuation theory and of quantum limiting behavior is not
restricted to nonequilibrium physics. The difference between equilibrium and nonequilibrium
macroscopic states lies mainly in the choice of macroscopic constraints. The constraints
describing equilibrium (like energy, particle number) usually commute and hence the problem we
discuss here typically
falls in a nonequilibrium context.\\

A first attack on this problem dates back to John von Neumann,
\cite{vN}.  His idea went as follows: The single particle position
and momentum operators $Q$, $P$ satisfy the commutation relation\footnote{Set the Planck constant
to one.} $[Q,P] = i$. Hence, assigning copies $Q_i$, $P_i$, $i = 1,\ldots,N$ to each of
$N$ particles, the averages $Q^N = \frac{1}{N}\sum_i Q_i$, $P^N =
\frac{1}{N}\sum_i P_i$ satisfy $[Q^N,P^N] = \frac{i}{N}$. Although
they do not commute and hence cannot be diagonalized together (nor simultaneously measured), one
can think of suitable modifications
$\tilde Q^N$, $\tilde P^N$ that already commute and that in a
sense well approximate the originals, at least for large $N$. Indeed, von Neumann explicitly
constructs commuting operators
$\tilde Q^N$, $\tilde P^N$ which have purely discrete spectra of
nondegenerate eigenvalues $(q^N_\al,p^N_\al)_\al$, and whose eigenvectors $(\psi^N_\al)_\al$ make
a complete orthonormal basis system in $L^2(\bbR)$. They approximate the operators $Q^N,P^N$ in
the sense that
\begin{align}
  (\psi^N_\al, Q^N \psi^N_\al) &= q^N_\al\,,\qquad
  (\psi^N_\al, P^N \psi^N_\al) = p^N_\al
\\ \intertext{and}
  \|(Q^N - q^N_\al)\, \psi^N_\al\| &\leq \frac{C}{\sqrt{N}}\,,\qquad
  \|(P^N - p^N_\al)\, \psi^N_\al\| \leq \frac{C}{\sqrt{N}}
\end{align}
with $C \approx 60$, see~\cite{vN} for details.\\

What follows is a slight modification and generalization of the above idea that comes close to the
point of view of quantum information theory. Instead of modifying the operators themselves, we
look for the largest or typical subspaces that in some sense well approximate the eigenspaces for
given eigenvalues, simultaneously for \emph{all} macroscopic observables from a collection. This
construction proves to be natural since the corresponding entropy, measuring the dimension of that
typical subspace, actually satisfies a variational principle. Hence, it can be directly compared
with another, more familiar although physically less satisfactory construction based on maximizing
the von Neumann entropy. The (non)equality of both entropies is then a problem of (non)equivalence
of ensembles. Looked at from another angle, such an equivalence gives a counting interpretation to
the von Neumann entropy in the thermodynamic limit, and opens interesting possibilities towards a
consistent and meaningful scheme of quantum large deviations.

\subsection{Quantum macrostates and entropy}\label{sec: macrostates}

A macroscopically large quantum system is modeled by a sequence of finite-dimensional Hilbert
spaces $(\caH^N)_{N\uparrow +\infty}$ on which we have standard traces $\Tr^N$. As macroscopic
observables we consider a collection $M^N = (M^N_k)_{k\in I}$ of self-adjoint operators on
$\caH^N$; for simplicity, we assume $I$ to be finite.
For each operator there is a projection-valued measure $\caQ^N_k$ on $\bbR$ such that, by the
spectral theorem,
\begin{equation}\label{eq: spectral}
  F(M^N_k) = \int_\bbR \caQ^N_k(\id z)\,F(z)\,,\qquad F \in C(\bbR)
\end{equation}
(which is just to say that $M^N_k$ is unitarily equivalent to a multiplication, or simply that
$M^N_k$ can be diagonalized.) A quantum counterpart of the classical set of microstates
$M^N_k \stackrel{\de}{=} m_k$ for some macrostate $m_k \in \bbR$ is the projection
\begin{equation}\label{eq: macrostate-q}
  \caQ^{N,\de}_k(m_k) = \int_{m_k-\de}^{m_k+\de} \caQ^N_k(\id z)
\end{equation}

\subsubsection{Commuting observables}

As a warm-up, assume first that $M^N$ is a collection of mutually commuting operators. In that
case,
$\caQ^N(\id z) = \prod_{k\in I} \caQ^N_k(\id z_k)$ is a common projection-valued measure,
\eqref{eq: spectral} extends to
\begin{equation}
  F(M^N) = \int_{\bbR^I} \caQ^N(\id z)\,F(z)\,,\qquad
  F \in C(\bbR^I)
\end{equation}
and a macrostate $m = (m_k)_{k\in I}$ gets represented by the projection
\begin{equation}
  \caQ^{N,\de}(m) = \prod_{k\in I} \caQ^{N,\de}_k(m_k)
\end{equation}
The classical entropy, say in the form~\eqref{eq: entropy-finite}, extends to
\begin{equation}
  S^{N,\de}(m) = \log \Tr^N[\caQ^{N,\de}(m)]
\end{equation}
This is a formalism entirely equivalent to the one for classical systems.

\subsubsection{General observables}\label{eq: quantum-general}

For a general collection $M^N$ of observables and a macrostate $m \in \bbR^I$, projections
$(\caP^N)_{N\uparrow +\infty}$ are said to be \emph{concentrating} at $m$,
written $\caP^N \to m$, whenever
\begin{equation}\label{eq: concentration}
  \lim_{N\uparrow +\infty} \tr^N[F(M^N_k) \rel \caP^N] = F(m_k)
\end{equation}
is satisfied for all $F\in C(\bbR)$ and $k \in I$, with the notation
\begin{equation}\label{eq: mc-ensemble}
  \tr^N[\,\cdot \rel \caP^N] = \frac{\Tr^N[\caP^N \cdot\,]}{\Tr^N[\caP^N]}
\end{equation}
for the normalized trace on $\caP^N\,\caH^N$. Condition~\eqref{eq: concentration} is a law of
large numbers for observables
$M^N_k$ under quantum state~\eqref{eq: mc-ensemble}; it can be equivalently written as the
condition
\begin{equation}\label{eq: concentration1}
  \lim_{N\uparrow +\infty} \tr^N[\caQ^{N,\de}_k(m_k) \rel \caP^N] = 1
\end{equation}
for any $\de > 0$ and $k\in I$. Physically, it means that all $M^N_k$, $k\in I$ are asymptotically
dispersionless under state~\eqref{eq: mc-ensemble}. \\

Having in mind the classical situation where the entropy counts the number of \emph{all}
microstates
$x$ such that $M^N(x) \stackrel{\de}{=} m$, we are mostly interested in those concentrating sequences
that are maximal in the sense of dimension counting. Hence, we
 define the (infinite-system,
Boltzmann) entropy
$s(m)$ for any $m = (m_k)_{k\in I}$ by the variational problem
\begin{equation}\label{eq: entropy-q}
  s(m) = \limsup_{\caP^N \to m} \frac{1}{N}\log \Tr^N[\caP^N]
\end{equation}
i.e., $s(m)$ is the largest limit point over all projections concentrating at $m$. Any projections
$\caP^N$ attaining the entropy $s(m)$ in the large
$N$ limit,
\begin{equation}
  \limsup_{N\uparrow +\infty} \frac{1}{N} \log \Tr^N[\caP^N] = s(m)
\end{equation}
are then called \emph{typical} projections concentrating at
$m$.\footnote{Note a slight difference in the terminology with
respect to\ \cite{DMN2} where the typical concentrating projections were rather called a
microcanonical macrostate. The present terminology is closer to the one of quantum information
theory.} Clearly, they provide a variant of the microcanonical ensemble for noncommuting
observables. An example comes in Section~\ref{sec: qKac}.

To check that the above definition of entropy is meaningful, we first revisit the relation between
the macroscopic autonomy and the H-theorem of Section~\ref{sec: H-general} in the present quantum
set-up. Second, we link our construction to the canonical construction based on maximizing the von
Neumann entropy, and we prove that they are equivalent under suitable conditions.

\subsection{H-theorem}

As a microscopic dynamics we consider a family of automorphisms\footnote{This means that
$\tau^N_t (X\,Y) = \tau^N_t(X)\,\tau^N(Y)$ for any $X,Y \in \caB(\caH^N)$, which is a
noncommutative generalization of classical deterministic map. Physically,
$(\caH^N,\tau^N_t,\Tr^N)$ models a \emph{closed} quantum dynamical system; note that
$\Tr^N(\tau^N_t(\cdot)) = \Tr^N(\cdot)$ and hence $\Tr^N$ corresponds to the invariant
(counting, unnormalized) classical measure.}
$(\tau^N_t)_{t \geq 0}$ acting on
the observables from $\caB(\caH^N)$ and having the semigroup property
\begin{equation}
  \tau^N_t\tau^N_s = \tau^N_{t+s}\,,\qquad t,s \geq 0
\end{equation}
Denote
\begin{equation}
  \Om_0 = \{m \in \bbR^I;\,s(m) \geq 0\}
\end{equation}
the set of all admissible macrostates. The conditions on the emergent macroscopic dynamics now
have the following form, cf.\ Section~\ref{sec: H-general}. There are maps
$(\phi_t)_{t\geq 0}$ on $\Om_0$ satisfying
\begin{enumerate}
\item
\emph{semigroup condition}: $\phi_t \circ \phi_s = \phi_{t+s}$, $t,s \geq 0$;
\item
\emph{autonomy condition}: for every $m \in \Om_0$ there exist some typical
projections $\caP^N \to m$ concentrating at $m$ such that for all
$F \in C(\bbR)$, $k \in I$, and $t \geq 0$,
\begin{equation}\label{eq: autonomy-q}
  \lim_{N\uparrow +\infty} \tr^N[\tau^N_t F(M^N_k) \rel \caP^N]
  = F((\phi_t m)_k)
\end{equation}
or, equivalently,
\begin{equation}\label{eq: autonomy-q1}
  \lim_{\de\downarrow 0} \lim_{N\uparrow +\infty} \tr^N[\tau^N_t \caQ^{N,\de}_k((\phi_t m)_k)
  \rel \caP^N] = 1
\end{equation}
\end{enumerate}
Under these condition it is easy to prove that
\begin{equation}
  s(\phi_t m) \geq s(\phi_s m)\,,\qquad m\in\Om_0,\,t \geq s\geq 0
\end{equation}

Indeed, let $\caP^N \to m$ be typical projections concentrating at $m$ and verifying~\eqref{eq:
autonomy-q} or \eqref{eq: autonomy-q1}. Using that
$\tau^N_t$ is invertible and $(\tau^N_t)^{-1}$ is again an automorphism such that
$\Tr^N((\tau^N_t)^{-1}\cdot) = \Tr^N(\cdot)$, one has
\begin{equation}
\begin{split}
  \tr^N[\tau^N_t F(M^N_k) \rel \caP^N]
  &= \frac{\tr^N[F(M^N_k)\,(\tau^N_t)^{-1} \caP^N]}{\tr^N[(\tau^N_t)^{-1} \caP^N]}
\\
  &= \tr^N[F(M^N_k) \rel (\tau^N_t)^{-1} \caP^N]
\end{split}
\end{equation}
Hence, autonomy~\eqref{eq: autonomy-q} implies that $(\tau^N_t)^{-1} \caP^N$ concentrate at
$\phi_t m$, $(\tau^N_t)^{-1} \caP^N \to \phi_t m$.\footnote{Note they do not have to be
\emph{typical} concentrating projections at $\phi_t m$!} As a result,
\begin{equation}
  s(m) = \limsup_{N\uparrow +\infty} \frac{1}{N}\log
  \Tr^N[(\tau^N_t)^{-1}\caP^N] \leq s(\phi_t m)
\end{equation}
By combining with the semigroup property,
\begin{equation}
  s(\phi_t m) = s(\phi_{t-s} \circ \phi_s m) \geq s(\phi_s m)
\end{equation}
as claimed.\\

Notice that in~\eqref{eq: autonomy-q}--\eqref{eq: autonomy-q1} we have required the autonomy
condition in the sense of a law of large numbers; compare with a much weaker assumption~\eqref{eq:
autonomy}. A possible way how to prove the H-theorem under a weaker autonomy condition here too,
might be via suitably weakening the notion of concentration and by modifying the definition of
entropy; we do not discuss this issue.

\subsection{Canonical formalism}

Von Neumann has introduced the entropy functional on states
$\om^N(\cdot) = \Tr^N[\rho^N\cdot\,]$ over $\caB(\caH^N)$ by
\begin{equation}
  \frS(\om^N) = -\Tr^N[\rho^N \log \rho^N]
\end{equation}
For trace states on subspaces of $\caH^N$, given as
$\om^N_{\caP^N}(\cdot) = \Tr^N[\caP^N\cdot\,] / \Tr^N[\caP^N]$, the von Neumann entropy boils down to
\begin{equation}
  \frS(\om^N_{\caP^N}) = \log \Tr^N[\caP^N]
\end{equation}
In this light, entropy~\eqref{eq: entropy-q} can also be written as
\begin{equation}\label{eq: entropy-q1}
  s(m) = \limsup_{\caP^N \to m} \frac{\frS(\om^N_{\caP^N})}{N}
\end{equation}

A general and very successful approach in statistical physics lies in the idea that a variational
principle like~\eqref{eq: entropy-q1} can often be extended to a larger `test space', so that (1)
a new variational problem becomes easier to solve, and (2) the resulting entropy $s(m)$ can be
proven to remain unchanged. This is a standard approach at least when describing thermal
equilibrium, but it is often used in a similar way to describe nonequilibrium macroscopic states
(sometimes then referred to as \emph{constrained} equilibria).\\

To obtain the \emph{canonical} description for a given macrostate $m \in \bbR^I$, we write
$\om^N \stackrel{1}{\to} m$ for any sequence of states satisfying
$\lim_{N\uparrow +\infty} \om^N(M^N_k) = m_k$, $k \in I$ (convergence in mean). Analogous
to~\eqref{eq: entropy-q1}, we define the \emph{canonical} entropy,
\begin{equation}
  s_\can(m) = \limsup_{\om^N \stackrel{1}{\to} m} \frac{\frS(\om^N)}{N}
\end{equation}
Any sequence of states $(\om^N)_{N\uparrow +\infty}$ such that
$\lim_{N\uparrow +\infty} \frS(\om^N)/N = s_\can(m)$ we then call \emph{canonical states} at
$m$.\\

An advantage of this formulation is that one can often find canonical states explicitly in a
Gibbsian form: consider states $\om^N_\la(\cdot) = \Tr^N[\rho^N_\la \cdot\,]$ defined as
\begin{equation}\label{eq: canonical states}
  \rho^N_\la = \frac{1}{\caZ^N_\la}\, e^{N \sum_k \la_k M^N_k}\,,\qquad
  \caZ^N_\la = \Tr^N[e^{N\sum_k \la_k M^N_k}]
\end{equation}
with some $\la = (\la_k)_{k\in I}$. If
$\lim_{N\uparrow +\infty} \om^N_\la(M^N_k) = m_k$, $k \in I$, then
$(\om^N_\la)_{N\uparrow +\infty}$
are canonical states at $m$.

This easily follows from the positivity of relative entropy, see e.g.~\cite{BR}: for any $\om^N
\stackrel{1}{\to} m$, $\om^N(\cdot) = \Tr^N[\rho^N \cdot\,]$,
\begin{equation}
\begin{split}
  \limsup_{N\uparrow +\infty} -\frac{1}{N}\, \om^N[\log \rho^N] &\leq
  \limsup_{N\uparrow +\infty} -\frac{1}{N}\,
  \om^N[\log \rho^N_\la]
\\
  &= \limsup_{N\uparrow +\infty} \frac{1}{N}
  \log\caZ^N_\la - \sum_k \la_k m_k
\\
  &= \limsup_{N\uparrow +\infty} -\frac{1}{N}\,
  \om^N_\la[\log \rho^N_\la]
\end{split}
\end{equation}
as claimed. It also yields the canonical entropy in the form
\begin{equation}\label{eq: can-explicit}
  s_\can(m) = p(\la) - \sum_k \la_k m_k
\end{equation}
where we have defined the `pressure'
\begin{equation}
  p(\la) = \limsup_{N\uparrow +\infty} \frac{1}{N} \log\caZ^N_\la
\end{equation}

\subsection{Macroscopic equivalence}\label{sec: equivalence}

By construction, $s_\can(m) \geq s(m)$. A natural question arises under what conditions both
entropies are actually equal. This is a familiar problem of the equivalence of ensembles
(microcanonical versus canonical in this case) on the level of entropies, however, the usual
arguments, e.g.~\cite{Li,Si,Ge}, are mostly restricted to the case of equilibrium and commuting
observables (with the energy and/or the particle number as the only variables). The generalized
microcanonical ensemble in the sense of Section~\ref{eq: quantum-general} requires some refinement
of those arguments. Below we provide some sufficient conditions
 for the equivalence.\\

Let $(\om^N_\la)_{N\uparrow +\infty}$ be canonical states~\eqref{eq: canonical states} with
$\om^N_\la \stackrel{1}{\to} m$. Assume that
\begin{enumerate}
\item
the limit
\begin{equation}
  p(\la) = \lim_{N\uparrow +\infty} \frac{1}{N} \log
  \Tr^N[e^{N\sum_k \la_k M^N_k}]
\end{equation}
exists and has the derivative
$\frac{\id p(\ka\la)}{\id\ka}\big|_{\ka = 1} = \sum_k \la_k\, m_k$;
\item
for any $j \in I$, the generating function
\begin{equation}
   q_j(\ka) = \lim_{N\uparrow +\infty} \frac{1}{N} \log
   \Tr^N[e^{N\sum_k \la_k M^N_k} e^{\ka N M^N_j}]
\end{equation}
exists and has the derivative
$\frac{\id q_j(\ka)}{\id\ka}\big|_{\ka = 0} = m_j$.
\end{enumerate}
Under these hypotheses we will prove that
\begin{equation}\label{eq: equivalence-main}
  s(m) = s_\can(m) = \sum_{k \in I} \la_k m_k - p(\la)
\end{equation}

Remark that by the Golden-Thompson inequality,\footnote{$e^{A+B} \leq e^A e^B$ for all hermitian
matrices $A,B$.}
\begin{equation}
  q_j(\ka) \geq p(\la + (0,\ldots,(\ka)_j,\ldots,0))
\end{equation}
Unless $M_N^k$ all mutually commute, this inequality generically becomes strict and those
$(q_j)_{j \in I}$ are fundamentally different from the pressure; they appear naturally
when studying quantum large fluctuations, see
Section~\ref{sec: qLD}.\\

The proof of equivalence~\eqref{eq: equivalence-main} comes in a sequence of steps: first we show
that the canonical states $\om^N_{\la}$ are exponentially concentrating at $m$, then we construct
typical projections for these states, and finally we prove that those typical projections
concentrate at $m$ too.

\subsubsection{Exponential concentration}\label{sec: q-exp}

By assumption, $q_j$ exists in some interval $[-\ka_0,\ka_0]$, $\ka_0 > 0$. From the spectral
theorem~\eqref{eq: spectral},
\begin{equation}\label{eq: ub-beginning}
\begin{split}
  q_j(\ka) &= p(\la) + \lim_{N\uparrow +\infty} \frac{1}{N}\log
  \om^N_{\la}[e^{\ka N M^N_j}]
\\
  &= p(\la) + \lim_{N\uparrow +\infty} \frac{1}{N}\log
  \int_\bbR \om^N_{\la}[\caQ^N_j(\id z)]\, e^{\ka z N}
\\
  &\equiv p(\la) + \lim_{N\uparrow +\infty} \frac{1}{N}\log
  \int_\bbR \nu^N_j(\id z)\, e^{\ka z N}
\end{split}
\end{equation}
where we have introduced the (classical) probability measures $\nu^N_j$ as the distribution of
$M^N_j$ under states $\om^N_{\la}$. Denote the last term as $\psi_j(\ka)$, and fix some $\de > 0$.
One has the estimate
\begin{equation}
  \int_\bbR \nu^N_j(\id z)\, e^{\ka z N} \geq e^{\ka(m_j + \de) N}
  \nu^N_j[z \geq m_j + \de]
\end{equation}
which implies, by the existence of the limiting generating function,
\begin{equation}
  \limsup_{N\uparrow +\infty} \frac{1}{N}\log
  \nu^N_j[z \geq m_j + \de] \leq \psi_j(\ka) - \ka(m_j + \de)
\end{equation}
for all $0 \leq \ka \leq \ka_0$. Since
$\frac{\id \psi_j}{\id\ka}|_{\ka = 0} = m_j$, there exists $\ka_1 = \ka_1(\de)$,
$0 < \ka_1 \leq \ka_0$ such that
$\psi_j(\ka_1) \leq \ka_1 m_j + \frac{\ka_1 \de}{2}$. Hence,
\begin{equation}
  \limsup_{N\uparrow +\infty}\frac{1}{N}\log \nu^N_j[z \geq m_j + \de]
  \leq -\frac{\ka_1\de}{2}
\end{equation}
Combining with an analogous argument for
$\nu^N_j[z \leq m_j - \de]$, we arrive at the bound
\begin{equation}\label{eq: ub-end}
   \om^N_{\la}[Q^{N,\de}_j(m_j)] \geq
   1 - e^{-C_j(\de) N}
\end{equation}
valid for all $\de > 0$ and $N \geq N_j(\de)$, with some $C_j(\de) > 0$ and $N_j(\de)$; the
$\caQ^{N,\de}_j(m_j)$ is given by~\eqref{eq: macrostate-q}.

This in particular implies that states $\om^N_{\la}$ are concentrating at $m(\la)$ in the sense of
a law of large numbers analogous to~\eqref{eq: concentration1}; moreover, the concentration is
exponentially fast. Note that the above argument is similar to the construction of large deviation
upper bounds, cf.~any textbook on the large deviation theory, e.g.\ \cite{DZ,dH}. \\

In an analogous way we exploit assumption (1) on the pressure. This time we consider the
observable $\sum_k \la_k M^N_k$ and denote by $\bar\caQ^N$ the corresponding projection-valued
measure, i.e., such that
\begin{equation}
  F\bigl(\sum_k \la_k M^N_k\bigr) = \int_\bbR \bar\caQ^N(\id z)\,F(z)\,,\qquad F\in C(\bbR)
\end{equation}
Repeating the arguments~\eqref{eq: ub-beginning}--\eqref{eq: ub-end}, we get the result
\begin{equation}\label{eq: pressure-ub}
   \om^N_{\la}[\bar Q^{N,\de}] \geq  1 - e^{-\bar C(\de) N}
\end{equation}
with
\begin{equation}
   \bar Q^{N,\de} = \int_\bbR \bar Q^N(\id z)\,
   \chi\bigl(\sum_k \la_k m_k -\de \leq z \leq \sum_k \la_k m_k +\de\bigr)
\end{equation}
valid again for all $\de > 0$, $N \geq \bar N(\de)$, with some $\bar C(\de) > 0$ and $\bar
N(\de)$.

\subsubsection{Typical projections}

From~\eqref{eq: pressure-ub} there is a sequence $\de_N \downarrow 0$ such that the projections
$\caP^N = \bar\caQ^{N,\de_N}$ satisfy
\begin{equation}\label{eq: typical sequence}
  \lim_{N\uparrow +\infty} \om^N_{\la}[\caP^N] = 1
\end{equation}
By construction one has the operator inequalities
\begin{equation}\label{eq: operator-ineq}
  \caP^N\bigl(\sum_k \la_k m_k - \de_N\bigr) \leq \caP^N \sum_k \la_k M^N_k \leq
  \caP^N\bigl(\sum_k \la_k m_k + \de_N\bigr)
\end{equation}
which yield the upper bound
\begin{align}
  \Tr^N[\caP^N] &= \om^N_\la[(\rho^N_\la)^{-1}\caP^N]
  \leq \caZ^N_\la e^{-N\bigl(\sum_k \la_k m_k - \de_N\bigr)}
  \om^N_\la[\caP^N]
\\\intertext{and the lower bound}\label{eq: q-lower}
  \Tr^N[\caP^N] &\geq \caZ^N_\la e^{-N\bigl(\sum_k \la_k m_k + \de_N\bigr)}
  \om^N_\la[\caP^N]
\end{align}
Using~\eqref{eq: typical sequence} and \eqref{eq: can-explicit}, this proves\footnote{Note that
one only needs that
$\lim_{N\uparrow +\infty} \frac{1}{N} \log \Tr^N[\caP^N] = 0$. In particular, the assumption on the
differentiability of the pressure is convenient but far from necessary; see also a comment below.}
\begin{equation}\label{eq: equivalence-proven}
  \lim_{N\uparrow +\infty} \frac{1}{N} \log\Tr^N[\caP^N] =
  p(\la) - \sum_k \la_k m_k = s_\can(m)
\end{equation}
As soon as we prove that projections $\caP^N$ are concentrating at $m$ (see the next section), the
last equation simply means that $s(m) \geq s_\can(m)$. Since the opposite inequality is obvious,
we arrive at~\eqref{eq: equivalence-main} as claimed.\\

The arguments used in this section are well known in both classical and quantum information
theory, and projections $\caP^N$ satisfying~\eqref{eq: typical sequence} and~\eqref{eq:
equivalence-proven} are usually called \emph{typical (sequence of) projections}. Their existence
under mild assumptions for a large class of models is a subject of the Shannon-McMillan(-Breiman)
theorem, see e.g.\ \cite{Bj2} and references therein. For a nice overview of the principles of
quantum information theory see~\cite{Da}.

\subsubsection{Concentration of typical projections}\label{sec: concentration-end}

To finish the proof we need to show that
 $\caP^N$ as constructed in the last section concentrate
at $m$. The following is true for any $Y^N \geq 0$:
\begin{equation}
\begin{split}
  \om^N_\la[Y^N] &= \Tr^N[(\rho^N_\la)^\frac{1}{2} Y^N (\rho^N_\la)^\frac{1}{2}]
\\
  &\geq \Tr^N[\caP^N (\rho^N_\la)^\frac{1}{2} Y^N (\rho^N_\la)^\frac{1}{2} \caP^N]
\\
  &= \Tr^N[(Y^N)^\frac{1}{2} \caP^N \rho^N_\la (Y^N)^\frac{1}{2}]
\\
  &\geq \frac{1}{\caZ^N_\la}\, e^{N\bigl(\sum_k \la_k m_k - \de_N\bigr)}
  \Tr^N[\caP^N]\, \tr^N[Y^N \rel \caP^N]
\\
  &\geq e^{-2 N \de_N} \om^N[\caP^N]\,\tr^N[Y^N \rel \caP^N]
\end{split}
\end{equation}
where we have used inequalities~\eqref{eq: operator-ineq} and
 \eqref{eq: q-lower}. Take now
$Y^N = 1 - \caQ^{N,\ep}_j(m_j)$ and use the
exponential concentration property of $\om^N_\la$, inequality~\eqref{eq: ub-end}; one obtains
\begin{equation}
  1 - \tr^N[\caQ^{N,\ep}_j(m_j) \rel \caP^N] \leq
  e^{-(C_j(\ep) - 2\de_N) N} (\om^N_\la[\caP^N])^{-1}
\end{equation}
for any $\ep > 0$ and $N \geq N_j(\ep)$, which immediately gives\footnote{Note it actually yields
an exponential concentration, even under that weaker assumption
$\lim_{N\uparrow +\infty} \frac{1}{N} \log \Tr^N[\caP^N] = 0$.}
\begin{equation}
  \lim_{N\uparrow +\infty} \tr^N[\caQ^{N,\ep}_j(m_j) \rel \caP^N] = 1
\end{equation}
Repeating for all $j \in I$, this proves $\caP^N \to m$.

\subsection{Towards quantum large deviations}\label{sec: qLD}

Using the notation of Section~\ref{sec: q-exp}, the G\"artner-Ellis theorem of (classical) large
deviations, \cite{DZ,dH}, teaches us that whenever the generating function $q_j \in C^1(\bbR)$ is
differentiable and strictly convex, one has the law
\begin{equation}\label{eq: qLD}
  \lim_{\de\downarrow 0} \lim_{N\uparrow\infty} \frac{1}{N}
  \log\nu^N_j[z \stackrel{\de}{=} \tilde m_j] = -I_j(\tilde m_j)
\end{equation}
for any $\tilde m_j$ such that
$\tilde m_j = \frac{\id q_j}{\id\ka}\big|_{\ka = \ka(\tilde m_j)}$ for some (unique by assumption)
$\ka(\tilde m_j)$,
and with the rate function $I_j$ being the Legendre transform
\begin{equation}\label{eq: q-rate}
\begin{split}
  I_j(\tilde m_j) &= \sup_\ka \bigl\{\ka\, \tilde m_j -
  \lim_{N\uparrow +\infty} \frac{1}{N}\log
  \om^N_\la\bigl[e^{\ka N M^N_j}\bigr] \bigr\}
\\
  &= \sup_\ka [\ka\, \tilde m_j - q_j(\ka) + p(\la)]
\\
  &= \ka(\tilde m_j)\,\tilde m_j - q_j(\ka(\tilde m_j)) + p(\la)
\end{split}
\end{equation}
(Naturally, for $m$ such that $\om^N_\la \stackrel{1}{\to} m$ one has
$\ka(m_j) = 0$ and $I_j(m_j) = 0$.) In terms of the canonical states
$\om^N_\la$, \eqref{eq: qLD} becomes simply
\begin{equation}\label{eq: qLD1}
  \lim_{\de\downarrow 0} \lim_{N\uparrow +\infty}
  \frac{1}{N}\log \om^N_\la[\caQ^{N,\de}_j(\tilde m_j)] = -I_j(\tilde m_j)
\end{equation}
This is an exponential law for the outcomes of measurements of the observables
$M^N_j$, upon the canonical states $\om^N_\la$. This gives an interpretation to $q_j$
as the corresponding generating function.

The existence and differentiability of $q_j$ gets nontrivial whenever the observables $M^N_k$ are
more complicated than just something like the spatial averages of one-site observables over a
lattice (a simple example are the observables $M^N_1, M^N_2, M^N_3$ in the quantum Kac model,
Section~\ref{sec: qKac}). In the usual context of quantum lattice models, no general argument is
known even for the existence of $q_j(\ka)$, which is in contrast to the case of pressure
$p(\la)$ where the situation is rather well understood, \cite{Si,Is,BR}.
For some partial results about the existence of $q_j$ in the so called high-temperature regime
see~\cite{NR,Rey}; the differentiability is studied in~\cite{NR}.\\

By means of the Varadhan lemma, \cite{DZ}, formula~\eqref{eq: qLD1} can be equivalently written as
\begin{equation}\label{eq: qLD2}
  \lim_{N\uparrow +\infty} \frac{1}{N}\log \om^N_\la [e^{N F(M^N_j)}]
  = \sup_z \{F(z) - I_j(z)\}
\end{equation}
for any $F \in C(\bbR)$, for simplicity assumed to be bounded from above. This form provokes still
another related question, namely the asymptotic limit
\begin{equation}\label{eq: qLD-PRV}
  \lim_{N\uparrow +\infty} \frac{1}{N}\log
  \Tr^N[e^{N\bigl(\sum_k \la_k M^N_k + F(M^N_j)\bigr)}]
\end{equation}
Although this is likely not directly related to the quantum fluctuations, such formulas appear
naturally when studying lattice models with a combination of short-range and long-range
interactions. Some authors consider \emph{this} formulation as a genuine problem of quantum large
deviations; see e.g.\ \cite{PRV} where the authors show the above limit to exist in the case of
$M^N_k$ being averages over one-site spin observables.\footnote{This in particular mean that the
canonical states $\om^N_\la$ are product states.} They prove the following variational principle:
\begin{align}
  \eqref{eq: qLD-PRV} &= \sup_z \{F(z) - I'_j(z)\}\,,\qquad
  I'_j(z) = \sup_\ka \{\ka\, z - q'_j(\ka)\}
\\
  q'_j(\ka) &= \lim_{N\uparrow +\infty} \frac{1}{N} \log
  \Tr^N\bigl[e^{N\bigl(\sum_k \la_k M^N_k + \ka M^N_j \bigr)}\bigr]
\end{align}
which is similar to~\eqref{eq: qLD}--\eqref{eq: q-rate} up to the modified generating function
$q'$.

A general and systematic quantum large deviation theory is lacking, however, and remains an
interesting open question. Possibly even more ambitious, both physically and mathematically, would
be the problem of correlated large fluctuations for noncommuting macroscopic observables. Some
ideas on this issue can be found in~\cite{Bj1}; also the present construction of generalized
microcanonical ensembles, Section~\ref{eq: quantum-general}, seems related to this problem.

\section{Example: quantum Kac ring}\label{sec: qKac}

This is a quantum extension of the Kac ring model of Section~\ref{sec: Kac}, introduced and
studied in~\cite{DF,DMN1}. Consider a ring $\La = \{1,\ldots,N\}$ again, and associate with each
site $i$ a quantum spin $\eta(i) \in \bbC^2$ and a classical variable $g(i) \in \{1,0\}$ that
indicates the presence respectively the absence of a scatterer. The state space of the model is
hence $\caH^N \times K^N$ with Hilbert space $\caH^N = \bbC^{2N}$ (spins) and classical space
$K^N = \{0,1\}^N$ (scatterers).

\subsection{Macroscopic description}\label{sec: qKac-macro}

As macroscopic observables we consider the operators
\begin{equation}
  M^N_\al = \frac{1}{N} \sum_{i=1}^N \si_\al(i)\,,\qquad
  \al = 1,2,3
\end{equation}
where $\si_\al(i)$ are copies at site $i$ of the Pauli matrices
\begin{equation}
  \si_1 = \left(
            \begin{array}{cc}
              0 & 1 \\
              1 & 0 \\
            \end{array}
          \right)\,,\quad
  \si_2 = \left(
            \begin{array}{cc}
              0 & -i \\
              i & 0 \\
            \end{array}
          \right)\,,\quad
  \si_3 = \left(
            \begin{array}{cc}
              1 & 0 \\
              0 & -1 \\
            \end{array}
          \right)
\end{equation}
representing three components of the local `magnetization', and
\begin{equation}
  M^N_0(g) = \frac{1}{N} \sum_{i=1}^N g(i)
\end{equation}
the density of scatterers. By construction,
$[M^N_1,M^N_2] = \frac{1}{N}M^N_3$ (and cyclic permutations).
The classical (= commutative) case is restored by keeping e.g.\
$M^N_0$ and $M^N_3$ as the only macroobservables.

It is sometimes convenient to embed $K^N$ in $\bbC^{2N}$ and to utilize a compact notation for
both operators on $\caH^N$ and classical functions on $K^N$. In this sense we speak below about
states over $\caH^N \times K^N$, and we use the shorthand
$\hat\Tr{}^N = \sum_{g \in K^N} \Tr^N$.\\

In the canonical framework,
\begin{equation}\label{eq: qKac-canonical state}
\begin{split}
  \om^N_\la(\cdot) &= \frac{1}{\caZ^N_\la}\,\hat\Tr{}^N[ e^{N\sum_{\al=0}^3 \la_\al M^N_\al} \cdot\,]
\\
  &= e^{-N p(\la)} \hat\Tr{}^N
  \bigl[e^{\sum_{i=1}^N \bigl(\la_0 g(i) + \sum_{\al=1}^3 \la_\al \si_\al(i) \bigr)} \cdot\,\bigr]
\end{split}
\end{equation}
are product canonical states, and the pressure is
\begin{equation}
  p(\la) = \frac{1}{N}\log \hat\Tr{}^N [e^{N\sum_{\al=0}^3 \la_\al M^N_\al}]
  = \log 2[(1 + e^{\la_0}) \cosh |\vec\la|]
\end{equation}
with the shorthands $\vec\la = (\la_1,\la_2,\la_3)$ and
$|\vec\la| = (\la_1^2 + \la_2^2 + \la_3^2)^\frac{1}{2}$. Further,
$\om^N_\la \stackrel{1}{\to} m$ where
\begin{equation}
  m_0 = \frac{\partial p}{\partial \la_0} = (1 + e^{-\la_0})^{-1}\,,\quad
  m_\al = \frac{\partial p}{\partial \la_\al} = \frac{\la_\al}{|\vec\la|}\tanh |\vec\la|\,,\
  \al = 1,2,3
\end{equation}
The canonical entropy is then, $m = (m_0,\vec m)$,
\begin{equation}\label{eq: qKac-entropy}
\begin{split}
  s_\can(m) &= p(\la) - \sum_{\al=0}^3 \la_\al m_\al
\\
&=
\begin{cases}
  - \frac{1+|\vec m|}{2} \log \frac{1+|\vec m|}{2}
  - \frac{1-|\vec m|}{2} \log \frac{1-|\vec m|}{2}\\
  \hspace{3mm} -m_0 \log m_0 - (1-m_0) \log (1-m_0)
  & \text{if } |\vec m| < 1,\, 0 < m_0 < 1
\\
  -\infty & \text{otherwise}
\end{cases}
\end{split}
\end{equation}
cf.~the classical case, \eqref{eq: Kac-entropy}.\\

To obtain a microcanonical description in the sense of Section~\ref{eq: quantum-general}, we
associate with any macroscopic state $m = (m_0,\vec m)$ the \emph{modified} macroscopic observable
$(M^N_0, \bar M^N)$,
\begin{equation}
  \bar M^N = \sum_{\al=1}^3 \frac{m_\al}{|\vec m|} M^N_\al
  = \frac{1}{N} \sum_{i=1}^N \bar\si(i)\,,\qquad
  \bar\si = \frac{\vec m}{|\vec m|} \cdot \vec\si
\end{equation}
and the modified macrostate $(m_0,|\vec m|)$.

Since $\bar\si$ is unitarily equivalent to e.g.\ $\si_3$, i.e.,\
$\bar\si = W \si_3 W^\dagger$ with some $W^\dagger = W^{-1}$, we
are back at the classical (commutative) situation. Denoting by
$\bar\caQ^N(\id z)$ the projection-valued measure for $(M^N_0,\bar
M^N)$, one easily checks that any\footnote{The notation is the same as in Section~\ref{sec:
macrostates}.}
$\bar\caQ^{N,\de_N}(m_0,|\vec m|)$ such that $\de_N \downarrow 0$,
are concentrating projections at $(m_0,|\vec m|)$. Moreover, if
$N^\frac{1}{2} \de_N \uparrow +\infty$ then these are
\emph{typical} concentrating projections at $(m_0,|\vec m|)$ and
the entropy is, as essentially can be read off from the classical formula~\eqref{eq: Kac-entropy},
\begin{equation}
  s(m_0,|\vec m|) = \lim_{N\uparrow +\infty} \frac{1}{N} \log
  \hat\Tr{}^N[\bar\caQ^{N,\de_N}(m_0,|\vec m|)] = s_\can(m_0,\vec m)
\end{equation}

In the last step, we need to show that $\bar\caQ^{N,\de_N}(m_0,|\vec m|)$ are also concentrating
at $m = (m_0,\vec m)$, that is the macrostate under the \emph{original} (noncommuting family of)
macroscopic observables $M^N$. This can be proven by essentially repeating the argument of
Section~\ref{sec: concentration-end}; we leave it to reader as an exercise. As a result, those
$\bar\caQ^{N,\de_N}(m_0,|\vec m|)$ are typical projections
concentrating at $m = (m_0,\vec m)$.

\subsection{Microscopic dynamics}

To model the scattering of quantum spins (represented
 by vector $\eta$) on the binary
variable $g$, consider a unitary matrix $V$ on $\bbC^2$,
\begin{equation}\label{eq: scatter}
  V = e^{i \vec h \cdot \vec\si}\,,\qquad \vec h = (h_1,h_2,h_3)
\end{equation}
Let the dynamics on $\caH^N \times K^N$ be given as, cf.~\eqref{eq: Kac-micro},
\begin{multline}
  U^N(\eta; g) = \bigl( g(N)\, V \eta(N) + (1-g(N))\,\eta(N),
\\
  g(1)\, V \eta(1) + (1-g(1))\, \eta(1),\ldots,
\\
  \ldots, g(N-1)\, V \eta(N-1) + (1-g(N-1))\,\eta(N-1);\,g \bigr)
\end{multline}
extended to a unitary operator in the quantum sector by linearity. The associated automorphisms
are then
\begin{equation}
  \tau^N_t(\cdot) = (U^N)^{-t} \cdot (U^N)^t
\end{equation}

\subsection{Macroscopic dynamics}

Let us start with a heuristic argument in the spirit of Boltzmann's Stosszahlansatz. Any
macrostate
$m = (m_0,\vec m)$ can be associated with the quantum state of a \emph{single} `effective' quantum
spin, via the $2 \times 2$ density matrix
\begin{equation}\label{eq: Bloch}
  \nu = \frac{1}{2}(\opunit + \vec m \cdot \vec\si)\,,\qquad
  \Tr[\nu\, \si_\al] = m_\al\,,\ \al = 1,2,3
\end{equation}
Each time step the effective spin either meets a scatterer (with probability $m_0$) or not (with
probability $1-m_0$).  Hence, its evolution is presumably $\nu \mapsto \nu_t = \bar\phi{}^t(\nu)$,
\begin{equation}
  \bar\phi(\nu) = m_0 V \nu V^\dagger + (1-m_0)\, \nu
\end{equation}
by construction enjoying the semigroup property. Using~\eqref{eq: scatter} and
\eqref{eq: Bloch}, this can be explicitly written as the evolution on macrostates:
$m_{t+1} = \phi(m_t)$ where
\begin{equation}\label{eq: qKac-evolution}
  \phi(m_0,\vec m) = (m_0,\ \vec m - 2m_0[(\vec n \times \vec m) \sin |\vec h| \cos |\vec h|
  -\vec n \times (\vec n \times \vec m) \sin^2 |\vec h|])
\end{equation}
with the notation $\vec n = \vec h / |\vec h|$. One easily checks that $\vec h \cdot \vec m$ is
invariant under
$\phi$; the evolution can be visualized as a spiral motion in the plane perpendicular to $\vec n$.
Provided that $|\vec h| \neq 0,\pi,2\pi,\ldots$ and $m_0 \in (0,1)$,
\begin{equation}
  \lim_{t \uparrow +\infty} \phi^t(m_0,\vec m) = (m_0,(\vec n \cdot \vec m)\,\vec n)
\end{equation}
and the relaxation is exponentially fast. The monotonicity of the entropy
$s(m_t)$ can also be easily verified.\\

A rigorous argument showing that the above heuristics is indeed correct was given in~\cite{qKac},
employing a strategy similar to that for the classical Kac model, Section~\ref{eq:
Kac-macroevolution}. The result reads that for a large class of (sequences of) states $\om^N
\stackrel{1}{\to} m$, including in particular
\begin{itemize}
\item
the (microcanonical) states $\tr^N[\,\cdot \rel \bar\caQ^{N,\de_N}(m,|\vec m|)]$ under those
typical concentrating projections at $m$ constructed in Section~\ref{sec: qKac-macro};
\item
the canonical states $\om^N_\la$ from~\eqref{eq: qKac-canonical state};
\end{itemize}
one has the law of large numbers:
\begin{equation}\label{eq: qKac-autonomy}
  \om^N[\tau^N_t F(M^N_\al)] = F((\phi_t m)_\al)\,,\qquad \al=1,2,3
\end{equation}
for all $F \in C(\bbR)$ and with $\phi$ given by~\eqref{eq: qKac-evolution}. Hence, one verifies
the autonomy condition~\eqref{eq: autonomy-q}.

\subsection{Exercise}

Consider $(M^N_0,M^N_3)$ as a new macroscopic observable. Check that the data $(m_0,m_3)$ are
macroscopically equivalent with
$(m_0,0,0,m_3)$ for the original `full' macroscopic observable
$M^N$; therefore the autonomy just follows from~\eqref{eq:
qKac-autonomy}. Calculate the entropy $s(m_0,m_3)$ and show that it oscillates as a function of
time. How can this apparent failure of the H-theorem be explained?

\section{Concluding remarks}

The text has discussed some newer and some older issues of nonequilibrium physics.  Main emphasis
has been on fluctuations and on the relation between entropy and irreversibility.  One could say
that everything has been an exploration of the idea that entropy production is a measure of
irreversibility.  Some central identities have been \eqref{ldb}, \eqref{jare}, \eqref{eq:
main-macro} and \eqref{resrea} which all point to the deep connection between source terms of
time-reversal breaking and statistical thermodynamic quantities.  They go beyond standard
irreversible thermodynamics because fluctuations play an essential role here.  As known since
long, the deviations of thermodynamic behavior are important in the very understanding of its
microscopic origin. These relations go also beyond the standard schemes as they are not
perturbative and they do not require linear approximations or
closeness to equilibrium.\\

 Nevertheless there is also a sense in
which all that has been attempted here does remain very close to the standard perspective.  We do
not mean only that there is not really much fundamentally new since Boltzmann's statistical
interpretation of entropy.  It is true that progress has been very slow and we have been writing
mostly from the point of view of the rear-guard, dissecting arguments and explanations that have
been won long before.  What we do have in mind however is that the theory so far remains very much
restricted to direct comparisons with equilibrium.  The obsession with time has mostly been an
interest in the passing away of structure, of deleting memory and of ending in equilibrium---all
the time centering around the second law of thermodynamics, and often applying Markovian schemes
or justifying molecular chaos. We hope that the lectures that are summarized in the preceding
sections have indeed clarified some of these issues, but we do not want to leave the reader
without
trying to provoke some feeling of totally different directions.\\

The most sensational instances of nonequilibrium physics are probably not be found in the problem
of relaxation to equilibrium nor in the installation of nonequilibrium via standard thermodynamic
forces for which the linear response theory appears to be working well even quite far from
equilibrium.   What needs to be understood is the constructive role of fluctuations far away from
equilibrium. For example, understanding nonequilibrium aspects in life processes be it for
molecular motors or for the problem of protein folding, requires fundamental studies in
reaction-rate theory. Ratchet mechanisms and the physics of transport and dissipation on very
small scales must be part of it also.  Nonequilibrium issues that are related to macroscopic
structure (even on cosmic scales), to pattern formation and to the organization of robust steady
behavior are mind-boggling, but one has to open them also via the methods and the traditions of
mathematical statistical physics when one wants its role to go further than ``simplification and
reduction of the results of previous investigations to a form in which the mind can grasp
them.''\footnote{J.~C.~Maxwell, in: {\it On Faraday's lines of forces.}}

\vspace{15mm}
\noindent{\bf Acknowledgment}\\
K.~N.~ acknowledges the support from the project AVOZ10100520 in
the Academy of Sciences of the Czech Republic.\\
The work of B.M.S. has been supported by a short term post doctoral grant at the Institute of
Theoretical Physics, K.U.Leuven; Grant of K.U.Leuven -- PDM/06/116 and Grant of Georgian National
Science Foundation -- GNSF/ST06/4-098.


\newpage
\bibliographystyle{plain}

\end{document}